\documentclass[%
superscriptaddress,
nofootinbib,
10pt,
 twocolumn,
 notitlepage,
 amsmath,amssymb,
 aps,
]{revtex4-1}

\usepackage{graphicx}
\usepackage{dcolumn}
\usepackage{bm}
\usepackage{color, colortbl}
\usepackage{comment}
\usepackage[textsize=small]{todonotes}
\usepackage[utf8]{inputenc}
\usepackage{placeins}	
\usepackage{feynmf}
\usepackage{array}      
\usepackage{todonotes}
\usepackage{slashed}
\usepackage{braket}
\usepackage{hyperref}

\renewcommand{\Re}{\ensuremath{\textrm{Re }}}
\renewcommand{\Im}{\ensuremath{\textrm{Im }}}
\newcommand{\abs}[1]{\ensuremath{\left|#1\right|}}
\renewcommand{\vec}[1]{\bm{#1}}

\newcommand{\sigmavec}{\vec{\sigma}}
\newcommand{\epsilonvec}{\vec{\epsilon}}

\makeatletter
\renewcommand*\env@matrix[1][*\c@MaxMatrixCols c]{%
  \hskip -\arraycolsep
  \let\@ifnextchar\new@ifnextchar
  \array{#1}}
\makeatother

\def\injannes(#1,#2,#3,#4,#5,#6){\begin{pmatrix}[cc|c]#1 & #2 & #3 \\ #4 & #5 & #6 \end{pmatrix}}

\newcommand{\diffcsdt}{\ensuremath{\frac{\textrm{d} \sigma}{\textrm{d} t}}}

\begin{document}
\title{Finite-Energy Sum Rules in Eta Photoproduction off the Nucleon}
\author{J.~Nys}
\email{Jannes.Nys@UGent.be}
\affiliation{Department of Physics and Astronomy, Ghent University, B-9000 Ghent, Belgium
}%
\author{V.~Mathieu}
\affiliation{Center for Exploration of Energy and Matter, Indiana University, Bloomington, IN 47403}
\affiliation{Physics Department, Indiana University, Bloomington, IN 47405, USA}

\author{C.~Fern\'andez-Ram\'{\i}rez}
\affiliation{Instituto de Ciencias Nucleares, Universidad Nacional Aut\'onoma de M\'exico, Ciudad de M\'exico 04510, Mexico}

\author{A.~N.~Hiller Blin}
\affiliation{Departamento de F\'isica Te\'orica and IFIC, Centro Mixto Universidad de Valencia-CSIC,
Institutos de Investigaci\'on de Paterna, E-46071 Valencia, Spain}
\affiliation{Center for Exploration of Energy and Matter, Indiana University, Bloomington, IN 47403}
\author{A.~Jackura}
\affiliation{Center for Exploration of Energy and Matter, Indiana University, Bloomington, IN 47403}
\affiliation{Physics Department, Indiana University, Bloomington, IN 47405, USA}
\author{M.~ Mikhasenko}
\affiliation{Universit\"at Bonn, Helmholtz-Institut f\"ur Strahlen- und Kernphysik, 53115 Bonn, Germany}
\author{A.~Pilloni}
\affiliation{Theory Center, Thomas Jefferson National Accelerator Facility,
Newport News, VA 23606, USA}
\author{A.~P.~Szczepaniak}
\affiliation{Center for Exploration of Energy and Matter, Indiana University, Bloomington, IN 47403}
\affiliation{Physics Department, Indiana University, Bloomington, IN 47405, USA}
\affiliation{Theory Center, Thomas Jefferson National Accelerator Facility,
Newport News, VA 23606, USA}
\author{G.~Fox}
\affiliation{School of Informatics and Computing, Indiana University, Bloomington, IN 47405, USA} 
\collaboration{Joint Physics Analysis Center}
\author{J.~Ryckebusch}
\affiliation{Department of Physics and Astronomy, Ghent University, B-9000 Ghent, Belgium
}%

\preprint{JLAB-THY-16-2384}

\begin{abstract}
The reaction $\gamma N \rightarrow \eta N$ is studied in the high-energy regime  (with photon lab energies \mbox{$E_\gamma^{\textrm{lab}} > 4 \,\text{GeV}$}) using information from the resonance region through the use of finite-energy sum rules (FESR). We illustrate how analyticity allows one to map the $t$-dependence of the unknown Regge residue functions. We provide predictions for the energy dependence of the beam asymmetry at high energies.
\end{abstract}

\maketitle
\section{Introduction}
Pseudoscalar-meson photoproduction on the nucleon is of current interest for hadron reaction studies. At low energies it provides information about the nucleon spectrum~\cite{Anisovich:2012ct,Kamano:2013iva,Ronchen:2015vfa,Chiang:2001as,Shklyar:2012js,Ruic:2011wf,He:2010ii} while at high energies it reveals details of the residual hadron interactions due to cross-channel particle (Reggeon) exchanges~\cite{Irving:1977ea}. These two regimes are analytically connected, a feature that can be used to relate properties of resonances in the direct channel to Reggeons in the cross channels. In practice this can be accomplished through dispersion relations and finite-energy sum rules (FESR)~\cite{Dolen:1967jr}.

In the resonance region there is abundant data on $\eta$ photoproduction on both proton and deuteron targets including polarization measurements~(see for example Refs.~\cite{Senderovich:2015lek,Akondi:2014ttg,Werthmuller:2014thb,Sumihama:2009gf,Crede:2009zzb,Fantini:2008zz}). On the other hand, higher energies ($E_{\text{lab}}>4$ GeV), only the unpolarized differential cross section has been measured~\cite{Braunschweig:1970jb,Dewire:1972kk}, providing little constrain on theoretical models. However, this is about to change thanks to the forthcoming data from the GlueX experiment at Jefferson Lab~\cite{Shepherd:2009zz,Ghoul:2015ifw}. 

Even though photons couple to both isospin $I=0,1$ states, there are some notable differences between high energy photoproduction of the $\eta$ ($I=0$) and the $\pi^0$ ($I=1$). The neutral pion differential cross section has a dip in the momentum transfer range, $-t \sim 0.5-0.6 \mbox{ GeV}^2$, whereas the $\eta$ meson differential cross section is rather smooth there. The dip in neutral pion photoproduction is likely to be associated with zeros in the residues of the two dominant Regge exchanges, the $\rho$ and the $\omega$~\cite{Guidal:1997hy,Yu:2011zu,Mathieu:2015eia}. It is an open question, however, what mechanisms are responsible for  filling in the dip in eta photoproduction. It is often assumed that large unnatural contributions come into play~\cite{Gault:1971dq,Worden:1972dc,Barker:1977pm,Sibirtsev:2010yj}. Finite-energy sum rules can provide clues here by relating the  $t$-dependence of  Regge amplitudes to that of the low-energy amplitude, usually described in terms of a finite number of partial waves. Early attempts could not resolve this issue due to the low quality of the data and the large uncertainties in the parametrization of the partial waves~\cite{Worden:1972dc,Barker:1977pm}.  Nowadays, however, there are several models that have been developed for the resonance region of  $\eta$ photoproduction~\cite{Chiang:2001as,Kamano:2013iva,Anisovich:2012ct,Ronchen:2015vfa, He:2010ii} allowing for a more precise FESR analysis.
Our main objective is to settle the discussion on the dip mechanism by invoking information from the low-energy regime. To this end, a Regge-pole model is fitted to the available high-energy cross-section data and compared to low-energy models through FESR. 
This work on $\eta$ photoproduction and ongoing work on $\pi^0$ photoproduction~\cite{vincent_fesr_pion} will set the stage for a combined low- and high-energy analysis of related reactions. 

As we discuss in this paper, the largest uncertainty in $\eta$ photoproduction stems from the unnatural parity Regge exchanges that in principle can be isolated through the photon beam asymmetry measurement. Such measurement will soon be published by the GlueX collaboration. The experiment uses linearly polarized photons with energy $E_\gamma^\textrm{lab} \sim 9$ GeV and it has simultaneously measured $\eta$ and $\pi^0$ production. This novel high-energy data will help to reduce the systematic uncertainties and to provide a better constrain on  Regge amplitudes for these reactions. 
Through the FESR analysis of $\eta$ photoproduction we make new  predictions based on the hypothesis of Regge-pole dominance to be compared with the forthcoming result from GlueX.

This  paper is organized as follows. In Section~\ref{sec:formalism} we discuss the formalism and set up all conventions with further details given in the Appendices.  Central to Regge theory, the topic of factorization is introduced in Section~\ref{sec:helicity_amps_factorization}. Section~\ref{sec:dispersion_relations} focuses on the  dispersion relation  used in the derivation of the FESR. The latter is presented in Section~\ref{sec:fesr}. In Section~\ref{sec:continuation} we  present the method used to analytically continue the low-energy amplitudes below the physical threshold which is needed in the calculation of the dispersive integral. The predictions arising from the low-energy side of the FESR, {\it i.e.} left-hand side (LHS) of the sum rules, are discussed in Section~\ref{sec:lhs} and compared to the high-energy data and the  Regge model in Section~\ref{sec:rhs}. The interpretation of the results and further development of the Regge model, in which we discuss possible contributions from the enigmatic  $\rho_2$ and $\omega_2$ exchanges, is given in Section~\ref{sec:exploratory}. Our conclusions are summarized in Section~\ref{sec:summary}.


\section{Formalism: Scalar Amplitudes}\label{sec:formalism}
We describe the kinematics of $\eta$ photoproduction on a nucleon target, the $s$-channel reaction, 
\begin{equation}\label{eq:schannel}
\gamma(k,\mu_\gamma) + N(p_i,\mu_i) \longrightarrow \eta(q) + N'(p_f,\mu_f) \,,
\end{equation}
by specifying particle four-momenta and helicities.
 We use $M_N$ and $\mu$ to denote the nucleon and $\eta$ masses, respectively. For all other particles we denote their masses by $m_x$.
Throughout this paper we use the standard Mandelstam variables
\begin{equation}
s = (k + p_i)^2 \,, \quad t=(k-q)^2 \,, \quad u = (k-p_f)^2 \,,
\end{equation}
related by $s+t+u= \Sigma = 2M_N^2 + \mu^2$. We refer to Appendix~\ref{sec:kinematics} for further details on the kinematics. The $u$ channel, in which the variable $u$ represents the physical center-of-mass energy squared of the $\gamma \overline{N} \rightarrow \eta \overline{N}$ reaction, is related to the $s$-channel by charge conjugation. To make this symmetry explicit, we use the crossing variable
\begin{equation}
\nu = \frac{s-u}{4 M_N} = \frac{s}{2M_N} + \frac{t-\Sigma}{4 M_N} = E_{\text{lab}} + \frac{t-\mu^2}{4 M_N}\,.
\end{equation}
Hence, the $t$ channel corresponds to $\gamma \eta \rightarrow \overline{N} N$. 
In order to formulate the dispersion relations, it is necessary to isolate and remove kinematical singularities. For this reason, it is convenient to work with the invariant amplitudes that are kinematic singularity free functions of the Mandelstam invariants. These amplitudes multiply four independent covariant tensors that contain the kinematical singularities. The tensor basis is constructed by combining the photon polarization vector $\epsilon^\mu \equiv \epsilon^\mu(k,\mu_\gamma)$  and particle momenta~\cite{CGLNpaper},
\begin{align}
M_1 &= \frac{1}{2} \gamma_5 \label{eq:M_1}
 \gamma_\mu \gamma_\nu F^{\mu \nu} \,, \\
M_2 &= 2  \gamma_5 q_\mu P_\nu F^{\mu \nu} \,, \\
M_3 &= \gamma_5 \gamma_\mu q_\nu F^{\mu \nu} \,, \\
M_4 &= \frac{i}{2} \epsilon_{\alpha \beta \mu \nu} \gamma^\alpha q^\beta F^{\mu \nu} \label{eq:M_4}\,.
 \end{align}
Here $P = (p_i + p_f)/{2}$ and $F^{\mu \nu} = \epsilon^\mu k^\nu - k^\mu \epsilon^\nu$. In terms of these 
 covariants the $s$-channel amplitude is given by  
\begin{align}\label{eq:helicity_amplitudes_definition}
A_{\mu_f, \mu_i \, \mu_\gamma} = \overline{u}_{\mu_f}(p_f) \left( \sum\limits_{k = 1}^4 A_k M_k(\mu_\gamma) \right) u_{\mu_i}(p_i) \,,
\end{align}
where the $A_k$ stand for the kinematic singularity and zero free amplitudes which contain the dynamical information on resonances and Regge exchanges.
It is convenient to decompose the invariant amplitudes in terms of amplitudes with well-defined isospin in the $t$-channel, $A^s$ and $A^v$ for $I=0$ and $I=1$, respectively,
\begin{equation}
A_i^{ab} = A^s_i \delta^{ab} + A^v_i \tau_3^{ab} \,,
\end{equation}
where $a$ and $b$ are the isospin indices of the two nucleons. Hence, 
\begin{subequations}
\begin{align}
A^p_i &=  A_i(\gamma p \rightarrow \eta p) = A^s_i + A^v_i \,,\\
A^n_i &= A_i(\gamma n \rightarrow \eta n) = A^s_i - A^v_i \,.
\end{align}
\end{subequations}
We will use the collective notation $A^\sigma_i$ for the isospin components ($\sigma = s,v$). 
For isoscalar, {\it e.g.} $\eta$ meson photoproduction, the $s$ and $u$ channel correspond to fixed $I=1/2$. 
It follows from the symmetry properties of the covariants $M_i$ under  
 $s \leftrightarrow u$ crossing  that the amplitudes $A^\sigma_i$ with $i = 1,2,4$ ($i=3$) are even (odd) functions of $\nu$, {\it i.e.}
\begin{align}
A^\sigma_i(-\nu - i \epsilon,t) = \xi_i A^\sigma_i(\nu + i \epsilon,t) \,, 
\end{align}
with $\xi_1 = \xi_2 = -\xi_3 = \xi_4 = 1$ and $\epsilon > 0$. The $t$-channel quantum numbers of the invariant amplitudes can be identified by projecting onto the $t$-channel parity-conserving helicity amplitudes. The latter can be decomposed in terms of the $L-S$ basis allowing for identification of the spin and parity (see Ref.~\cite{Mathieu:2015eia} and references therein). For $\gamma N \rightarrow \eta N$, we list the invariant amplitudes in Table~\ref{tab:contributions_to_Ai} together with the corresponding quantum numbers and possible $t$-channel exchanges. We note that the amplitude $A_2' = A_1 +tA_2$, instead of $A_2$, has good $t$-channel quantum numbers~\cite{Mathieu:2015eia}. We will work with the set of amplitudes $\left( A_1, A_2', A_3, A_4 \right)$ which  allow to separate natural from unnatural parity $t$-channel contributions. The $\gamma \eta$ state couples to $C=-1$ exchanges in the $t$-channel, which for the $N\overline{N}$ state implies $C=(-1)^{L+S} = -1$. For the $N \overline{N}$ state, parity is given by $P=(-1)^{L+1}$.
Thus, for positive parity the total angular momentum is odd ($J=L$), while for negative parity, $J$ is either odd or even ($J=L\pm 1, L$). Furthermore, since $C = -1$ the $N \overline{N}$ state has $G$-parity equal to $-1$ for $I = 0$ and $+1$ for $I=1$.
Beside known resonances, $t$-channel exchanges with $J^{PC}=(2,4,...)^{--}$ are also allowed, but no mesons with these quantum numbers have been clearly observed\footnote{There are some experimental indications of the existence of $\rho_2$ and $\omega_2$ mesons~\cite{Anisovich:2002su,Anisovich:2011sva}. However, these states are  observed  by  a  single  group and poorly established and thus need confirmation~\cite{reviewofparticlephysics}.} to date~\cite{reviewofparticlephysics}. These quantum numbers are not exotic (only the $0^{--}$ is) and both the quark model and lattice QCD results predict the existence of such states~\cite{Godfrey:1985xj,Dudek:2013yja}. At high energies the dominant $t$-channel contributions in $\eta$ photoproduction are expected from the natural exchanges, which according to Table~\ref{tab:contributions_to_Ai} feed into the $A_1$ and $A_4$ amplitudes. The $C$-parity conservation  prohibits exchanges of the signature partners of the $\rho$ and $\omega$, the $a_2(1320)$ and $f_2(1270)$, respectively.
The amplitudes for isovector exchanges ($\rho$, $b$ and $\rho_2$) on proton and neutron 
 differ by sign. Schematically, the net contribution of $t$-channel exchanges considered here is given by 

\begin{align}
\gamma p \rightarrow \eta p\,, \qquad & A = (\omega + h + \omega_2) + (\rho+b+\rho_2) \,,\\
\gamma n \rightarrow \eta n\,, \qquad & A = (\omega + h + \omega_2) - (\rho+b+\rho_2) \,.
\end{align}

\begin{table}[tbh]
\def\arraystretch{2.}
\centering
\caption{Invariant amplitudes $A_i$ with corresponding $t$-channel exchanges. $I$ is isospin, $G$ is $G$-parity, $J$ is total spin, $P$ is parity, $C$ is charge conjugation, $\eta=P(-1)^J $ is the naturality. \label{tab:contributions_to_Ai} }
\begin{tabular}{>{$}c<{$}|>{$}c<{$}|>{$}c<{$}|>{$}c<{$}|>{$}c<{$}}
\bm{A_i}	& \bm{I^G}	& \bm{J^{PC}}					& \bm{\eta}	& \textrm{\textbf{Leading exchanges}} \\
\hline\hline
A_1 & 0^-, 1^+ 	& (1,3,5,...)^{--}	& +1		& \rho(770), \omega(782)	\\
A_2' & 0^-, 1^+ 	& (1,3,5,...)^{+-}	& -1		& h_1(1170), b_1(1235)	\\
A_3 & 0^-, 1^+ 	& (2,4,...)^{--}		& -1		& \rho_2 (??), \omega_2 (??)	\\
A_4 & 0^-, 1^+ 	& (1,3,5,...)^{--}	& +1		& \rho(770), \omega(782)
\end{tabular}
\end{table}

At large $s$ the expression for the differential cross section and the photon beam asymmetry ($\Sigma$) simplifies and in terms of the scalar amplitudes is given by
\begin{align}
\diffcsdt &= \frac{1}{32 \pi} \left( \abs{A_1}^2 - t \abs{A_4}^2 + \abs{A_2'}^2 - t \abs{A_3}^2\right) \,, \label{eq:diffcs_expression_leading_s} \\
\Sigma \diffcsdt &= \frac{1}{32 \pi} \left( \abs{A_1}^2 - t \abs{A_4}^2 - \abs{A_2'}^2 + t \abs{A_3}^2\right) \,,
\end{align}
while the exact expression for the differential cross section reads
\begin{align}
\diffcsdt = \frac{1}{64 \pi s \abs{\vec{k}}^2}\frac{1}{2} 
\sum\limits_{\mu_f,\mu_i = \pm} \abs{A_{\mu_f,\mu_i\,1}}^2 \,,
\end{align}
since negative photon helicities are related by parity conservation.

\section{Tests of factorization}\label{sec:helicity_amps_factorization}
One of the main purposes of this paper is to investigate whether the high-energy data can be described entirely in terms of factorizable Regge poles~\cite{Gribov:1962fw}, or other contributions are needed. Specifically we investigate the implications of angular-momentum conservation which gives a stronger constraint on Regge amplitudes as compared to its implications for the scattering amplitude  in general. In the $s\to \infty$ limit, $s$-channel angular-momentum conservation  implies that the $s$-channel helicity amplitudes in Eq.~\eqref{eq:helicity_amplitudes_definition} have the following behavior as $t \to 0$ (see Appendix~\ref{sec:factorization})
\begin{align}\label{eq:t_behaviour_angular_momentum}
A_{\mu_f,\mu_i\,\mu_\gamma} \underset{t \rightarrow 0}{\sim} (-t)^{n/2} \,,
\end{align}
where $n = \abs{(\mu_\gamma - \mu_i) - (-\mu_f)} \geq 0$ is the net $s$-channel helicity flip. This is a weaker condition  than the one imposed by angular-momentum conservation on factorizable Regge amplitudes, 
 \begin{align}\label{eq:t_behaviour_factorization}
A_{\mu_f,\mu_i\,\mu_\gamma} \underset{t \rightarrow 0}{\sim} (-t)^{(n+x)/2} \,,
\end{align}
where $n+x=\abs{\mu_\gamma} + \abs{\mu_i - \mu_f} \geq 1$. 
We summarize the expected behavior for the four independent helicity amplitudes in Table~\ref{tab:t_behaviour}. It can be seen that when factorization is imposed, all helicity amplitudes in the Regge-pole model vanish at $t = 0$. If only the condition given in Eq.~\eqref{eq:t_behaviour_angular_momentum} is imposed, the  $s$-channel nucleon helicity flip amplitude  $A_{-,+\,1}$ can be finite at $t = 0$.

At leading order in $s$, and for small $\abs{t}$, the $s$-channel helicity amplitudes are related to the invariants, $A_i$ by~\cite{Mathieu:2015eia,Worden:1972dc}
\begin{align}
\frac{1}{\sqrt{2} s} \left( A_{+, + \, 1} + A_{-, - \, 1} \right) =& \sqrt{-t} A_4 \label{eq:contains_no_n_0_a}\\
\frac{1}{\sqrt{2} s } \left( A_{+, - \, 1} - A_{-, + \, 1} \right) =& A_1 \label{eq:contains_n_0_a}\\
\frac{1}{\sqrt{2} s } \left( A_{+, + \, 1} - A_{-, - \, 1} \right) =& \sqrt{-t} A_3 \label{eq:contains_no_n_0_b}\\
\frac{1}{\sqrt{2} s} \left( A_{+, - \, 1} + A_{-, + \, 1} \right) =& - A_2' = -(A_1 +t A_2) \label{eq:contains_n_0_b}
\end{align}
Thus, at high energies the invariants $A_3$ and $A_4$ ($A_1$ and $A_2'$) correspond to the $s$-channel nucleon-helicity non-flip (flip), respectively.  
Combining Eqs.~\eqref{eq:contains_n_0_a} and \eqref{eq:contains_n_0_b} we obtain
\begin{align}\label{eq:non_zero_hel_amp}
A_{-, + \, 1} = -\frac{s}{\sqrt{2}} \left(A_2' + A_1\right) \,.
\end{align}
We find that angular-momentum conservation does not require any of the invariant amplitudes $A_i$ to vanish at $t = 0$, but the stronger condition of Eq.~\eqref{eq:t_behaviour_factorization} implies that the Regge residues of  $A_1$ and $A_2'$ ought to vanish. 

The FESR test factorization by relating the $t=0$ behavior of the high-energy, Regge amplitudes to one at low energy, obtained for example from the phase-shift analysis.

\begin{table}[tbh]
\def\arraystretch{2.}
\centering
\caption{Behavior of the $s$-channel helicity amplitudes $A_{\mu_f, \mu_i \, 1}$ for given nucleon helicities, as predicted by Eqs.~\eqref{eq:t_behaviour_angular_momentum} and~\eqref{eq:t_behaviour_factorization}.\label{tab:t_behaviour}}
\begin{tabular}{c|c|c}
$\bm{A_{\mu_f, \mu_i \, 1}}$ 	& $\bm{n}$ & $\bm{n+x}$ \\
\hline\hline
$A_{-, - \, 1}$	& $1$	& $1$ \\
$A_{-, + \, 1}$	& $0$	& $2$ \\
$A_{+, - \, 1}$	& $2$	& $2$ \\
$A_{+, + \, 1}$	& $1$	& $1$
\end{tabular}
\end{table}

\section{Dispersion relations}\label{sec:dispersion_relations}
\begin{figure}[tbh]
\centering
\includegraphics[width=6cm]{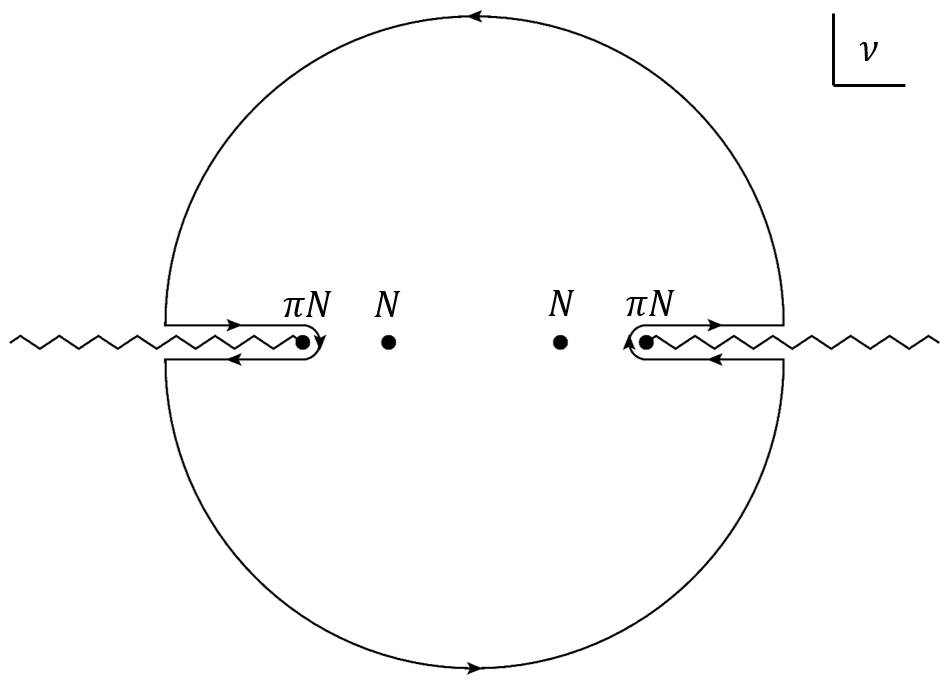}
\caption{Contour in the complex $\nu$ plane used for the dispersion relations. The $s$- and $u$-channel nucleon pole and $\pi N$ threshold and cut are shown.\label{fig:contour}}
\end{figure}

We assume that the scalar amplitudes have only the real axis dynamical cuts imposed by unitarity,  and we write the dispersion relations for  $A_i^\sigma(\nu,t)$ at constant $t$ using the contour in the $\nu$-plane shown in Fig.~\ref{fig:contour}. 

\begin{figure}[tbh]
\centering
\begin{minipage}{0.23\textwidth}
\includegraphics[width=\textwidth]{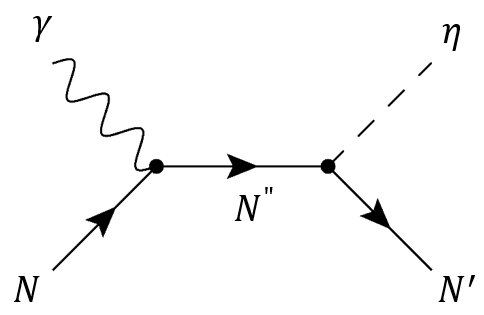}
\end{minipage}
\begin{minipage}{0.23\textwidth}
\includegraphics[width=\textwidth]{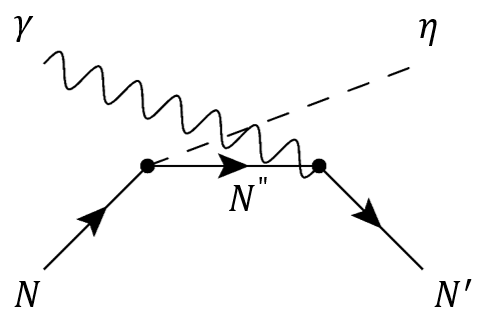}
\end{minipage}
\caption{Nucleon pole contributions to the dispersion relations.\label{fig:s_u_channel}}
\end{figure}

In Fig.~\ref{fig:contour} we identify the nucleon pole and a cut starting from the $\pi N$ threshold. 
We relate the residues of the $s$ and $u$ channel poles to the phenomenological couplings by identifying them with the Born terms calculated using an effective Lagrangian~\cite{Benmerrouche:1994uc} as shown in Fig.~\ref{fig:s_u_channel},
\begin{align}
\mathcal{L}_{\gamma N N} =& -e \overline{N} \gamma_\mu \frac{1 + \tau_3}{2} N A^\mu  \nonumber \\
&+ \frac{e}{4 M_N} \overline{N} (\kappa^s + \kappa^v \tau_3) \sigma_{\mu \nu} N F^{\mu \nu}\,,\\
\mathcal{L}_{\eta N N} =& -i\zeta g_{\eta NN}  \overline{N} \gamma_5 N \phi_\eta +  (1-\zeta) \frac{g_{\eta NN}}{2M_N} \overline{N} \gamma_\mu \gamma_5 N \partial^\mu \phi_\eta,
\end{align}
where $\kappa^s = \frac{1}{2}(\kappa_p + \kappa_n)$ and $\kappa^v = \frac{1}{2}(\kappa_p - \kappa_n)$ are the isoscalar and isovector nucleon
anomalous magnetic moments and $\sigma^{\mu \nu} = \frac{i}{2} [\gamma^\mu,\gamma^\nu]$. The two limiting cases are the $\zeta = 0$ pseudovector (PV) and $\zeta = 1$ pseudoscalar (PS) coupling. The role of these two couplings has been explored in dynamical models for the scattering amplitude based on effective Lagrangians~\cite{Tiator:1994et,Benmerrouche:1994uc,FernandezRamirez:2007fg}. In the Born terms, however, the difference between the two interactions leads to a non-pole contribution that does not contribute to the on-shell scattering amplitude for which the dispersion relation is written. The derivative term reduces
indeed to the other one upon use of the equation
of motion.

  
For the Born terms the two diagrams in Fig.~\ref{fig:s_u_channel} give (see Appendix~\ref{sec:nucleon_pole_contr}) 
\begin{align}
A_1^{\sigma,\textrm{pole}}(\nu,t) &= \frac{e g_{\eta NN}}{4 M_N} \left[ \frac{1}{\nu - \nu_N} - \frac{1}{\nu + \nu_N}\right] \,, \\
A_2^{\sigma,\textrm{pole}}(\nu,t) &= -\frac{e g_{\eta NN}}{4M_N^2} \left[ \frac{1}{(\nu - \nu_N)(\nu + \nu_N)}\right] \nonumber \\
&= -\frac{e g_{\eta NN}}{(t-\mu^2) 2M_N} \left[ \frac{1}{\nu - \nu_N} - \frac{1}{\nu + \nu_N}\right]\,, \\
A_3^{\sigma,\textrm{pole}}(\nu,t) &= -\frac{e g_{\eta NN}}{M_N} \frac{\kappa^\sigma}{4 M_N}\left[ \frac{1}{\nu - \nu_N} + \frac{1}{\nu + \nu_N}\right] \,, \\
A_4^{\sigma,\textrm{pole}}(\nu,t) &= -\frac{e g_{\eta NN}}{M_N} \frac{\kappa^\sigma}{4 M_N}\left[ \frac{1}{\nu - \nu_N} - \frac{1}{\nu + \nu_N}\right] \,,
\end{align}
where $\nu_N ={(t - \mu^2)}/{(4 M_N)}$.
The coupling $g_{\eta NN}$ is less known than $g_{\pi NN}$.  Using the latter and SU(3) symmetry one finds $g_{\eta NN}^2/{4\pi} = 0.9-1.8$ (where the uncertainty is induced by the uncertainty on the F/D ratio)~\cite{Benmerrouche:1994uc,Dumbrajs:1983jd}. 
On the other hand from fits to the $\eta$ photoproduction data using effective and chiral Lagrangian models~\cite{Tiator:1994et,FernandezRamirez:2007fg}, one obtains a smaller value, $g_{\eta NN}^2/{4\pi} = 0.4-0.52$. Similar results are found in the quark  models of  Refs.~\cite{Neumeier:2000fb,Kirchbach:1996kw}, while other constituent-quark models find an even smaller value, $g_{\eta NN}^2/{4\pi} = 0.04$~\cite{Saghai:2001yd}. In the following we choose $g_{\eta NN}^2/{4\pi}=0.4$ as a canonical value. On the real axis the dispersion relations for $(\xi_{i=1,2,4} = +1)$ are given by
\begin{align}\label{eq:dispersion_plus}
\Re A^\sigma_i(\nu,t) = B^{\sigma}_i(t) \frac{2 \nu_N}{\nu_N^2 - \nu^2} + \frac{2}{\pi} \mathcal{P}\int_{\nu_\pi}^{+\infty} \nu' \frac{\Im A^\sigma_i(\nu',t)}{\nu'^2 - \nu^2} d\nu' \,,
\end{align}
and for $(\xi_{i=3} = -1)$ by
\begin{align}\label{eq:dispersion_minus}
\Re A^\sigma_i(\nu,t) = B^{\sigma}_i(t) \frac{2 \nu}{\nu_N^2 - \nu^2} + \frac{2 \nu}{\pi} \mathcal{P}\int_{\nu_\pi}^{+\infty} \frac{\Im A^\sigma_i(\nu',t)}{\nu'^2 - \nu^2} d\nu' \,.
\end{align}
The residues $B^\sigma_i(t)$ of the nucleon poles are tabulated in Table~\ref{tab:pole_contributions_B}.

\begin{table}[tbh]
\def\arraystretch{2.}
\centering
\caption{Pole contributions to the dispersion relations in Eqs.~\eqref{eq:dispersion_plus}--\eqref{eq:dispersion_minus}.}
\label{tab:pole_contributions_B}
\begin{tabular}{l|c|c|c}
             & $\bm{\sigma = s}$                                             & $\bm{\sigma = v}$\\ 
\hline\hline
$B^\sigma_1$ & $\displaystyle{-\frac{e g_{\eta NN}}{4 M_N}}$                                 & $\displaystyle{-\frac{e g_{\eta NN}}{4 M_N}}$ &$e = 0.303$\\
$B^\sigma_2$ & $\displaystyle{\frac{e g_{\eta NN}}{2 M_N} \frac{1}{t-\mu^2}}$                & $\displaystyle{\frac{e g_{\eta NN}}{2 M_N} \frac{1}{t-\mu^2}}$ & $\displaystyle{g_{\eta NN}^2/{4\pi} = 0.4}$\\
$B^\sigma_3$ & $\displaystyle{\frac{e g_{\eta NN}}{4 M_N} \frac{\kappa^s}{M_N}}$ & $\displaystyle{\frac{e g_{\eta NN}}{4 M_N} \frac{\kappa^v}{M_N}}$ & $\kappa^s = -0.065$\\
$B^\sigma_4$ & $\displaystyle{\frac{e g_{\eta NN}}{4 M_N} \frac{\kappa^s}{M_N}}$ & $\displaystyle{\frac{e g_{\eta NN}}{4 M_N} \frac{\kappa^v}{M_N}}$ & $\kappa^v = 1.845$
\end{tabular}
\end{table}

\section{Finite-Energy Sum Rules}\label{sec:fesr}
For the high-energy part of the amplitude, we use a Regge parametrization. The contribution of a Regge pole with signature $\tau = (-1)^J$ to scalar amplitudes $A_i$ is given by~\cite{collins}
\begin{align}\label{eq:regge_R}
A_{i,R}(\nu,t) &= -\beta_i(t) \frac{\tau (r_i \nu)^{\alpha(t)} + (-r_i \nu)^{\alpha(t)}}{\sin \pi \alpha(t)} (r_i\nu)^{-1} \,, \\
&= -\beta_i(t) \frac{\tau + e^{ - i \pi \alpha(t)}}{\sin \pi \alpha(t)} (r_i\nu)^{\alpha(t) - 1} \label{eq:regge_R_2} \,,
\end{align}
where Eq.~\eqref{eq:regge_R_2} is the reduction on the real axis of the more general expression in Eq.~\eqref{eq:regge_R}. The $r_i$ is a scale parameter of dimension $\textrm{GeV}^{-1}$  and the residues $\beta_i(t)$ are dimensionless. 
Under crossing 
\begin{equation}
A_{i,R}(-\nu,t) = -\tau A_{i,R}(\nu,t)\,,
\end{equation}
where $\tau = -1$ ($+1$) for vector (tensor) exchanges. 
The $A_i \sim \nu^{\alpha - 1}$ behavior corresponds to the typical $\nu^{\alpha}$ behavior for the $s$-channel helicity amplitudes (see Eqs.~\eqref{eq:contains_no_n_0_a}--\eqref{eq:contains_n_0_b}). 
 Regge theory does not determine the residues $\beta (t)$ uniquely. They can be fixed, for example by comparing with the data. It follows from unitarity, however, that in the $s$-channel physical region, both $\beta(t)$ and $\alpha(t)$ are real.
The Regge amplitudes in Eq.~\eqref{eq:regge_R}, being analytical functions of $\nu$, can be represented via a dispersive integral,
\begin{widetext}
\begin{equation}\label{eq:dispersion_regge_part}
\Re A^\sigma_{i,R}(\nu,t) = \frac{1}{\pi} \mathcal{P}\int_0^{+\infty} \Im  A^\sigma_{i,R}(\nu',t) \left[\frac{1}{\nu' - \nu} + \frac{\xi_i}{\nu' + \nu}\right] d\nu' \,.
\end{equation}
\end{widetext}
If, for a particular energy $\Lambda$, the scalar amplitudes $A^\sigma_i$ 
can be approximated by the Regge form $A^\sigma_i(\nu,t) = A^\sigma_{i,R}(\nu,t)$ for $\nu > \Lambda$, then  Eqs.~\eqref{eq:dispersion_plus},~\eqref{eq:dispersion_minus} and~\eqref{eq:dispersion_regge_part} lead to the FESR~\cite{Mathieu:2015gxa}, 
\begin{align}\label{eq:FESR_main}
\frac{\pi B^\sigma_i(t)}{\Lambda} \left(\frac{\nu_N}{\Lambda}\right)^k +& \int_{\nu_\pi}^\Lambda \Im A^\sigma_i(\nu',t) \left(\frac{\nu'}{\Lambda}\right)^k \frac{d\nu'}{\Lambda} \nonumber \\
=& \beta^\sigma_i(t) \frac{\left(r_i\Lambda\right)^{\alpha(t)-1}}{\alpha(t)+k} \,,
\end{align}
which are used for even (odd) integer $k$ corresponding to $\xi_i = -1$ ($\xi_i = 1$), respectively. The energy $\Lambda$ denotes the transition energy between the low- and high-energy regime.
In order to derive Eq.~\eqref{eq:FESR_main}, one expands the combination of Eqs.~\eqref{eq:dispersion_plus},~\eqref{eq:dispersion_minus} and~\eqref{eq:dispersion_regge_part} in powers of $\nu'/\nu < 1$ (since $\nu' < \Lambda$ and $\nu > \Lambda$), after which the result follows from the condition for the coefficients of $(1/\nu)^{k}$. Hence, in principle Eq.~\eqref{eq:FESR_main} is satisfied for all even (odd) integer $k$ for each crossing odd (even) invariant amplitude. Alternatively, one can derive Continuous-Moment Sum Rules (CMSR) which also require the real part of the low-energy amplitude~\cite{OlssonCMSR,FerroFontan:1972in}.
The LHS of the FESR is a function of $t$ determined by the low-energy behavior of the scattering amplitude. The right hand side (RHS) is a function of $t$ determined by the high-energy behavior,  which we parametrize by  Regge poles.
 Amplitude zeros  or other features of the $t$-dependence seen on the LHS side will be linked to the residue functions $\beta(t)$. 

\section{Subthreshold continuation}\label{sec:continuation}
\begin{figure}[bth]
\centering
\includegraphics[width=0.4\textwidth]{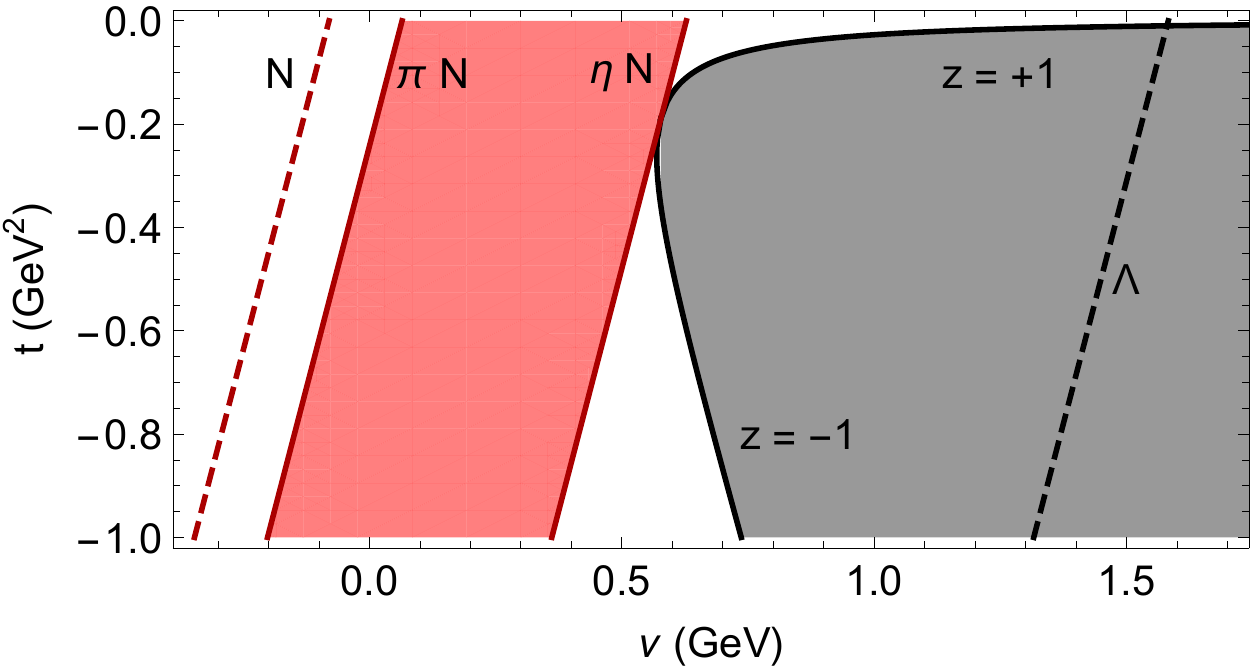}
\caption{Overview of the kinematic domains. The red band is the unphysical subthreshold region where we continue the amplitudes. The solid red lines indicate the $\pi N$ and $\eta N$ branch points at $\sqrt{s} = 1.07 \mbox{ GeV}$ and $1.49 \mbox{ GeV}$, respectively. The red dashed line shows the nucleon pole. The solid black lines show the boundaries $t(z_s\equiv\cos \theta = \pm 1)$ of the physical domain, which is indicated by the dark shade. The white domain between the $\eta N$ and physical boundary lines shows the unphysical domain above threshold where we use the multipole expansion to reconstruct the amplitudes. The black dashed line shows the upper boundary $\Lambda$ of the low-energy dispersion integral in Eq.~\eqref{eq:FESR_main}.\label{fig:regions_of_singularities}}
\end{figure}

The integral on the LHS of the FESR of Eq.~\eqref{eq:FESR_main} starts at the lowest $s$-channel threshold, {\it i.e.} the $\pi N$ threshold, and it is necessary to analytically continue the scalar amplitudes below the physical $\eta N$ threshold (see Fig.~\ref{fig:regions_of_singularities}). 

\begin{figure*}[tbh]
\centering
\begin{minipage}{1.0\textwidth}
\includegraphics[width=\textwidth]{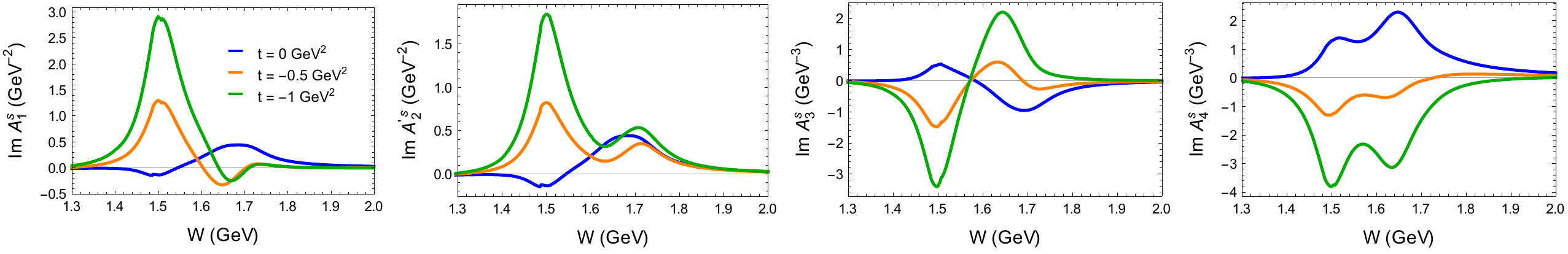}
\end{minipage}%

\begin{minipage}{1.0\textwidth}
\includegraphics[width=\textwidth]{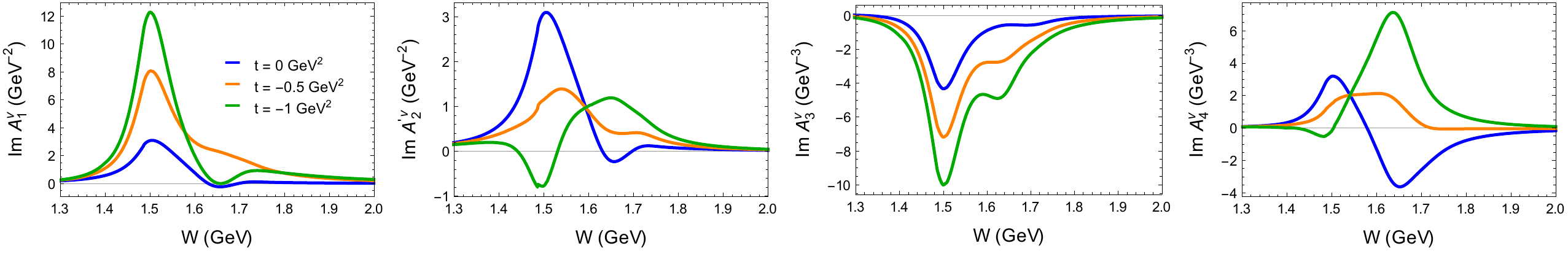}
\end{minipage}%
\caption{Isoscalar and isovector invariant amplitudes $(A_1,A_2',A_3,A_4)$ of the $\eta$-MAID 2001 model~\cite{Chiang:2001as} at $t = 0, -0.5$ and $-1 \textrm{ GeV}^2$.\label{fig:invariant_isospin}}
\end{figure*}

Low energy parametrizations that are currently available are based on the partial-wave series expansion. The series diverges in the unphysical domain and approximations are required. In the following we collectively denote the electric ($E_{l\pm}$) and magnetic ($M_{l\pm}$) multipoles, by $\mathcal{M}_{l\pm}$. Specifically, we use the $\eta$-MAID 2001~\cite{Chiang:2001as} model and an approach similar to Ref.~\cite{Aznauryan:2003zg} to continue the multipoles in $s=W^2$. For individual moments we identify resonances that give the dominant contributions close to threshold and continue them below threshold using the Breit-Wigner parametrization used in the $\eta$-MAID 2001 model~\cite{Chiang:2001as}. The formalism is summarized in Appendix~\ref{sec:eta_maid_formalism}.
At the $\eta N$ threshold, the multipoles behave as $\mathcal{M}_{l\pm} \underset{\abs{\vec{q}}\rightarrow 0}{\sim} \abs{\vec{q}}^l$, where $\vec{q}$ is the relative three-momentum in the $\eta N$ center-of-mass frame, and from  Eq.~\eqref{eq:costhetacm_and_q} it follows that  $\cos \theta \abs{\vec{q}}$ is linear in $t$ and finite at threshold (except on the boundary of the physical region where it is $0$). Thus, even though individual multipoles are suppressed at threshold they give a finite contribution at fixed $t$.  Some multipoles are dominated at threshold by a well-known resonance. For example, in the $\eta N$ photoproduction channel, the $E_{0+}$ is substantial at threshold in the physical region due to the $S_{11}(1535)$ resonance, which couples strongly to the $\eta N$ channel. But there are also multipoles where it is not clear how much they should contribute below the $\eta N$ threshold.
In practice, we identify the main multipole contributions to the invariant amplitudes at the $\eta N$ threshold and we continue them below threshold until no discontinuities are notable at threshold within the considered domain $0 \leq -t \leq 1 \mbox{ GeV}^2$. We hereby start from the lowest multipole order $l = 0$ and add subthreshold-continued higher-order multipoles until the invariant amplitudes below the $\eta N$ threshold (generated from a lower number of partial waves) sufficiently reproduce the amplitudes at threshold. 
The resulting isospin components of the continued invariant amplitudes are shown in Fig.~\ref{fig:invariant_isospin}. We note that the continuation becomes less reliable as $-t$ increases and we restrict the analysis to the range $0 \leq -t \leq 1\mbox{ GeV}^2$.

\section{Left-hand side of the FESR}\label{sec:lhs}
We proceed with the discussion of the LHS of the FESR~\eqref{eq:FESR_main}. Various features of the observed $t$-dependence will be analyzed in the context of the Regge parametrization in the following section. 

To compute the LHS we use a single parametrization for the low-energy amplitudes from the $\eta$-MAID 2001 model~\cite{Chiang:2001as}. Three main restrictions hinder the use of other available models. First, the sum rules in Eq.~\eqref{eq:FESR_main} require isospin decomposable amplitudes, meaning that a proton and neutron version of the low-energy model must be available. Second, the ingredients of the low-energy model should be simple enough and well tabulated in the corresponding references in order to allow for a reconstruction of the model. The latter is mandatory to enable a subthreshold continuation of the model amplitudes. For example, the Bonn-Gatchina model~\cite{Anisovich:2012ct} does provide a set of isospin decomposed multipoles. However, we were unable to continue the invariant amplitudes below the $\eta N$ threshold starting from the provided multipoles. A third restriction is that the low-energy models should be valid up to sufficiently high energies ($W \gtrsim 2 \mbox{ GeV}$).
A different version of the $\eta$-MAID model (dubbed $\eta$-MAID 2003) was presented in Refs.~\cite{Chiang:2002vq,Fix:2007st} with the aim to remedy the overestimated $D_{15}(1675)$ contribution in the $\eta$-MAID 2001 model. The model includes Regge contributions. However, since their parametrization is significantly different from the standard definition in Eq.~\eqref{eq:regge_R}, we do not include it in our analysis.

After carrying out the FESR analysis with the $\eta$-MAID 2001 amplitudes, we will compare the results to the Bonn-Gatchina 2014-02 (BoGn)~\cite{Anisovich:2012ct}, ANL-Osaka (ANL-O)~\cite{Kamano:2013iva} and Julich-Bonn (JuBo)~\cite{Ronchen:2015vfa} model for the proton target. For the latter two models, only the proton amplitudes are available. Furthermore, as discussed earlier, for all these other models, it is unclear how to extrapolate the invariant amplitudes outside the physical region $\abs{\cos \theta} \leq 1$.

\begin{figure*}[tbh]
\centering
\includegraphics[width=\textwidth]{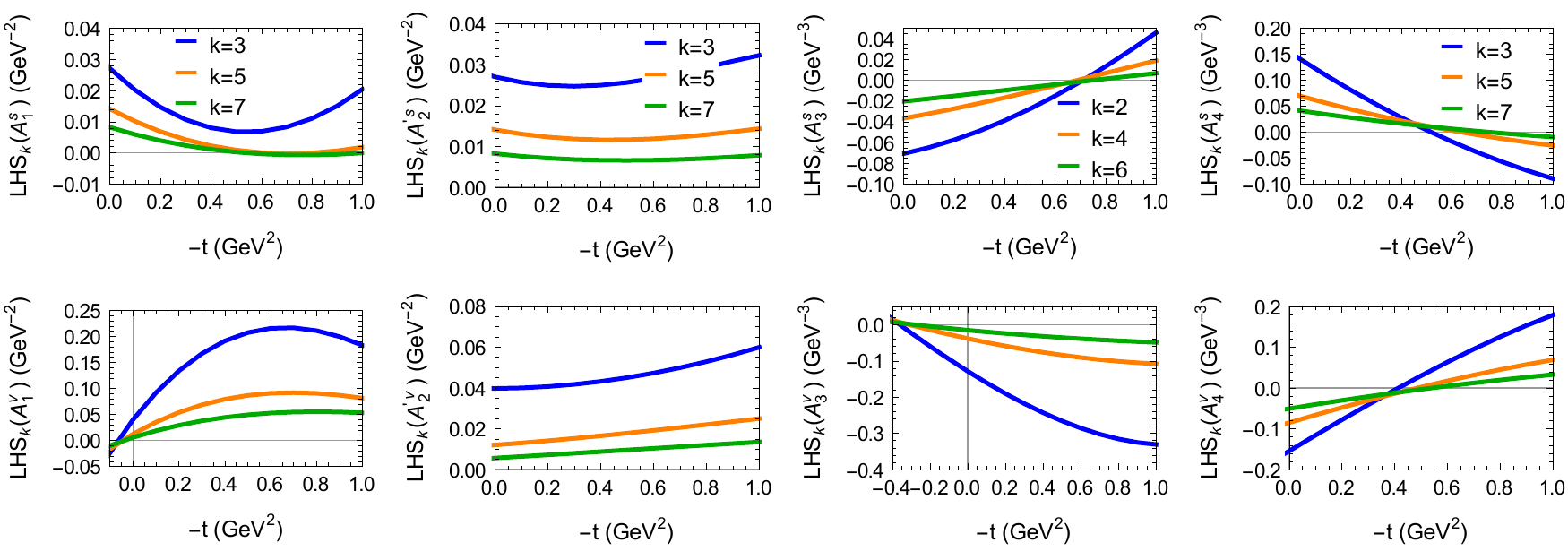}
\caption{LHS of the FESR Eqs.~\eqref{eq:dispersion_plus}--\eqref{eq:dispersion_minus}. We ignore the lowest moment, in order to soften the dependence on the unphysical subthreshold region. The results depicted on the top (bottom) four panels are for the isoscalar (isovector) component of the amplitude.\label{fig:LHS_fesr}}
\end{figure*}

\begin{figure*}[bth]
\centering
\includegraphics[width=\textwidth]{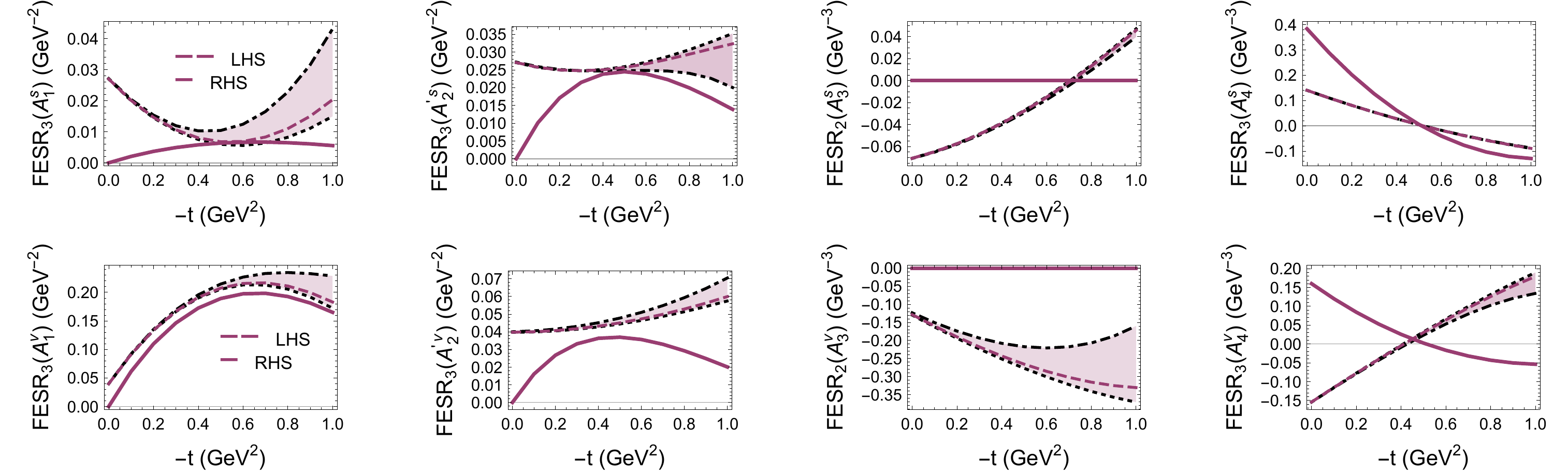}
\caption{Matching of the RHS with the LHS of the sum rules. The color band shows the range of the LHS, using a range of $0.04 \leq g_{\eta NN}^2/{4\pi} \leq 6.075$ couplings. The maximal (minimal) coupling is given by the dot-dashed (dotted) line. The coupling $g_{\eta NN}^2/{4 \pi} = 0.4$ corresponds to the dashed line. The RHS obtained from the Regge model is indicated by the solid line.\label{fig:lhs_vs_rhs_2}}
\end{figure*}

Fig.~\ref{fig:regions_of_singularities} illustrates the domain where the LHS of the FESR are evaluated and the different kinematic domains covered therein.  Note that the $s$- and $u$-channel $\pi N$ thresholds start to overlap at $\nu_\pi(t_\pi) = 0$ or $t_\pi = -0.243 \textrm{ GeV}^2$ (see Eq.~\eqref{eq:nu_pi}). 
In principle, at higher $-t$, the Schwarz reflection principle is no longer applicable, since the scattering amplitude is no longer real on a part of the real axis. From analyticity of the scattering amplitude in $t$, it is assumed that the dispersion relations can be applied beyond $-t_\pi$. The $\eta$-MAID 2001 model is applicable from threshold up to $W_{\textrm{max}} = 2 \textrm{ GeV}$ or $E_{\gamma,\textrm{max}}^{\textrm{lab}} = 1.66 \textrm{ GeV}$. Therefore, we are forced to take $\Lambda = E_{\gamma,\textrm{max}}^{\textrm{lab}} + (t-\mu^2)/4M_N$. 

\begin{figure}[bth]
\centering
\includegraphics[width=0.5\textwidth]{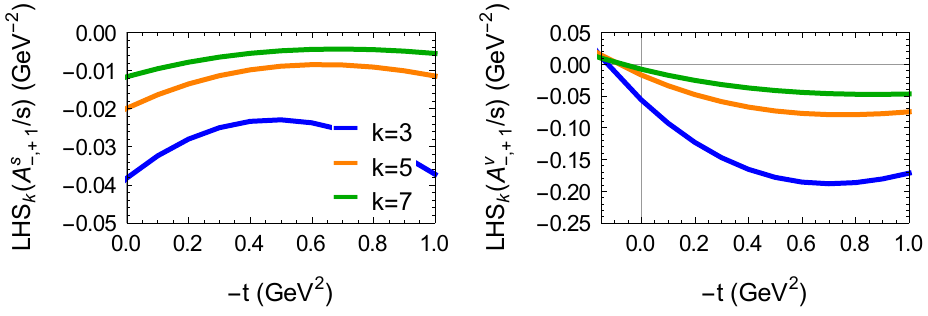}
\caption{LHS of the sum rules in Eq.~\eqref{eq:FESR_main} for the $s$-channel isoscalar and isovector contribution to the helicity amplitude in $A_{-,+\, 1}$ Eq.~\eqref{eq:non_zero_hel_amp}.\label{fig:non_zero_hel_amp_LHS}}
\end{figure}

The $\eta$-MAID 2001 model incorporates the nucleon Born terms, real $t$-channel $\rho$ and $\omega$ exchanges, and nucleon resonances up to the $F_{15}$ partial wave. Hence, the imaginary part of the model amplitudes can be reconstructed by including the  $l \leq 3$ multipoles. The results for the LHS of the FESR are shown in Fig.~\ref{fig:LHS_fesr}. Below we comment on the specific features observed in its $t$-dependence. We concentrate on moments with $k > 1$, since Eq.~\eqref{eq:FESR_main} assumes $\alpha + k > 0 $ and in order to reduce sensitivity to the subthreshold continuation. We also show the LHS of the FESR for a single moment in Fig.~\ref{fig:lhs_vs_rhs_2}, where the contribution of the Born terms is illustrated. It turns out that the main features (\textit{i.e.} relative strength and zeros) in the LHS can be attributed to the dispersive integral. Therefore, we discuss Fig.~\ref{fig:LHS_fesr} in terms of the dispersive term only. To facilitate the discussion on factorization, we also include the sum rules for the $s$-channel helicity amplitude $A_{-,+\,1}$ (see Eq.~\eqref{eq:non_zero_hel_amp}) in Fig.~\ref{fig:non_zero_hel_amp_LHS}.

\begin{itemize}

\item Comparing the LHS for the two isospin components of $A_1$, we find a dominant isovector contribution. Considering $A_1^{s,v}$ in Fig.~\ref{fig:invariant_isospin} at \textit{e.g.} $t = -0.5 \mbox{ GeV}^2$, the large $\mbox{LHS}(A_1^v)$ can be traced back to strong resonance contributions just above threshold and a smaller contribution at $W = 1.6-1.7 \mbox{ GeV}$ which both carry the same sign in $A_1^v$. While the bump around $W = 1.5$ GeV also dominates $A_1^s$, its isoscalar component is substantially smaller than its isovector part. Also, the second bump at $W = 1.6-1.7 \mbox{ GeV}$ enters the isoscalar amplitude with an opposite sign and  reduces the dispersive integral in $\mbox{LHS}(A_1^s)$. 
The large $\mbox{LHS}(A_1^v)$ is consistent with the expectation for the high-energy side of the sum rule, which is related 
 to a large $s$-channel nucleon-helicity flip component of the $t$-channel $\rho$ exchange. The small $\mbox{LHS}(A_1^s)$ is related to a negligible helicity-flip component of the $\omega$ (see Section~\ref{sec:rhs} for details).

\item The bump around $W = 1.5 \mbox{ GeV}$ in $A_1$ has a strong $t$-dependence due to its $D_{13}(1520)$ content. In both $A_1^s$ and $A_1^v$ the bump is smallest around $t = 0$. As a result, $\mbox{LHS}(A_1^v)$ tends towards zero for $t \to 0$. For the isoscalar component on the other hand, the smaller contribution at higher $W$ dominates the dispersive integral at $t = 0$, resulting in a different behavior of $\mbox{LHS}(A_1^s)$ as $t \to 0$. This second contribution to the $A_1$ is mainly attributed to the $D_{15}(1675)$ and $P_{11}(1710)$ within  the $\eta$-MAID 2001 model.

\item The LHS of the FESR for $A_4^s$ is large and switches sign at $t \approx -0.5 \mbox{ GeV}^2$. This behavior is generated in the low-energy model by the contributions around $W = 1.5$ and $1.65 \mbox{ GeV}$ which collectively switch sign at $t \approx -0.5 \mbox{ GeV}^2$. In the isovector component, these two contributions work destructively. However, the predictions for the LHS are quite similar to the isoscalar component, since its dispersive integral is dominated by the resonances around $W = 1.65$ GeV and hence, follows its sign switch. The opposite sign of the $\mbox{LHS}(A_4^s)$ and $\mbox{LHS}(A_4^v)$ is therefore mainly an effect induced by the contributions around $W = 1.65 \mbox{ GeV}$. The relative size of the $A_1^s$ and $A_4^s$ ($A_1^v$ and $A_4^v$) is related in the high-energy model to a dominant nucleon-helicity non-flip (flip) contribution of the $\omega$ ($\rho$).

\item The LHS of the $A_3^v$ FESR is quite substantial, which is a feature that is not expected from the perspective of the high-energy model. In fact, there are no known mesons which feed into the $A_3$ amplitude. Therefore, one would expect the sum rules for $A_3^{s,v}$ to be small compared to the other amplitudes. Considering Fig.~\ref{fig:invariant_isospin}, it appears that this contribution is mainly related to a bump around $W = 1.5 \mbox{ GeV}$ and to a smaller extent a constructively contributing peak around $W = 1.65 \mbox{ GeV}$. For the isoscalar part, the dominant peak around $W = 1.5 \mbox{ GeV}$ is substantially smaller than in the isovector component. On top of that, the second peak  contributes with an opposite sign, resulting in a smaller LHS for the isoscalar component of $A_3$. The two bump structures are mainly the result of the $D_{13}(1520)$ and $D_{15}(1675)$ resonance content. Since it is known~\cite{Chiang:2002vq} that the $D_{15}(1675)$ is overestimated within the $\eta$-MAID 2001 model, the non-negligible LHS predictions for $A_3^v$ might be a model-specific feature.

\end{itemize} 

We now focus on specific features (such as zeros) in the LHS of the FESR that will be used to constrain the high-energy model.

\begin{itemize} 
\item The LHS of the FESR for the amplitude $ A_1^v$, shows a zero at $t \approx 0.05 \textrm{ GeV}^2$. A zero at $t = 0$ is expected from Regge pole factorization. Indeed, it can be seen from Eqs.~\eqref{eq:contains_n_0_a} that $A_1$ must vanish at $t = 0$ for factorizable contributions, since all $s$-channel helicity amplitudes vanish.

\item 
To study the factorization properties, consider the LHS of the $A_{-,+\,1}^v$ FESR in Fig.~\ref{fig:non_zero_hel_amp_LHS}. This $s$-channel helicity amplitude was shown to be the only amplitude which is not forced to be zero at $t = 0$ by angular-momentum conservation (see Section~\ref{sec:helicity_amps_factorization}).
The tendency towards zero at $t = 0$ is not seen in the isoscalar component of the $A_{-,+1}$ amplitude which is a manifest violation of factorization. However, it should be noted that the $A^s_{-,+1}$ is small and might actually be consistent with zero at $t = 0$ within uncertainties of the model.  
The observed possible departure from factorization has also been seen in other reactions. A well-known example is charged pion photoproduction, where the factorization of the pion exchange term predicts a dip in the cross section at $t=0$, while the observed cross section is finite in the range $0 \leq -t \leq m_\pi^2$~\cite{Irving:1977ea}. In the latter case this may be attributed to the conspiring contribution from $s$-channel exchanges required by current conservation~\cite{Cho:1969qk}. Alternatively it may be due to absorption, whose effect on the amplitude can be taken approximately into account by evaluating the numerator of the pion exchange at $t = m_\pi^2$, also known as the Williams' “poor man absorption” model \cite{Williams:1970rg}.
\item For both the isovector and isoscalar component of the $A_4$, we observe a zero in the LHS of the FESR in the vicinity of $t \approx -0.5 \textrm{ GeV}^2$.
\item The low energy predictions of the FESR for $A_2^{\prime s}$ and $A_2^{\prime v}$ suggest a similar behavior with a relative strength $v/s \approx 1.5$. 
\end{itemize}

\begin{figure*}[htb]
\centering
\includegraphics[width=\textwidth]{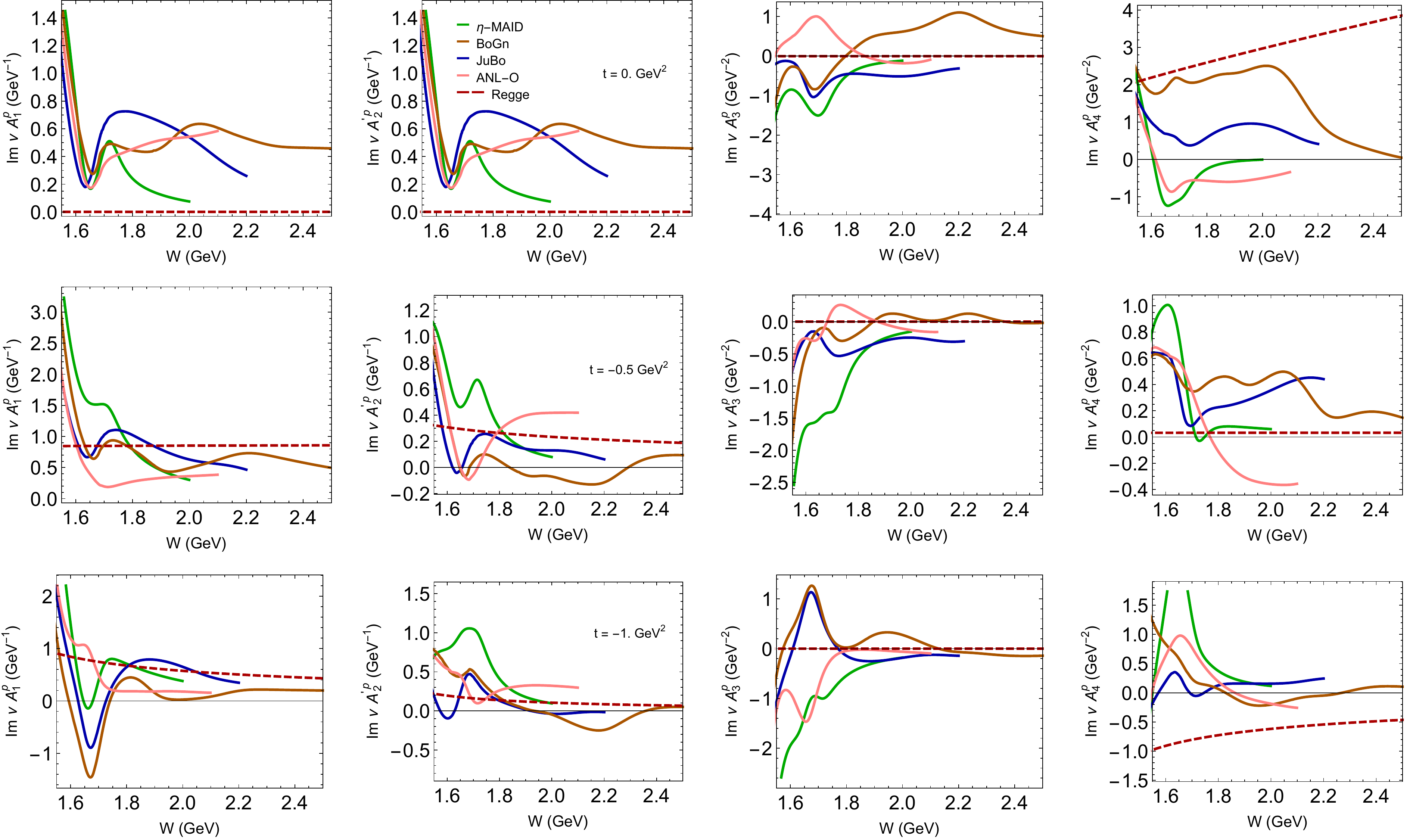}
\caption{Comparison of the proton-target amplitudes $A_i^p = A_i^s + A_i^v$ of the high-energy Regge prediction to a number of resonance-region models for $t=0, -0.5$ and $-1$ GeV${^2}$ from top to bottom. The models have been constructed using their $l\leq 5$ multipoles only.\label{fig:regge_comparison_to_models}}
\end{figure*}

More contemporary and coupled-channel models, such as Refs.~\cite{Kamano:2013iva,Anisovich:2012ct,Ronchen:2015vfa} might provide more decisive information on some of the above-mentioned observations. For example, consider $\Im A_{1}^p$ and $\Im A_{2}^{\prime p}$ at $t = 0$ in Fig.~\ref{fig:regge_comparison_to_models}. These models tend to predict a strong violation of factorization in the high-energy $\omega(/\rho)$ and $b/h$ exchanges compared to the somewhat older $\eta$-MAID 2001 model. Especially evaluating the $A_3$ FESR with state-of-the-art coupled channels might shed light on the unexpectedly large $A_3$ contribution. However, such an analysis is currently hindered by the problematic subtreshold region and low predictive power and instabilities just above threshold, just outside of the physical region.

\section{Right-hand side of the FESR}\label{sec:rhs}
The RHS of the FESR are evaluated using a Regge pole model. Inspired by the observations made in the previous section, one is able to determine the $t$ dependence of the Regge pole residues $\beta_i^\sigma(t)$ within the domain $0 \leq -t \leq 1 \mbox{ GeV}^2$. 
The most direct way of using the FESR is by computing the LHS of the FESR using a low-energy model, and extracting the residues (by inverting Eq.~\eqref{eq:FESR_main}) by introducing only assumptions about the Regge trajectories. However, directly implementing the low-energy predictions for the residues into a high-energy model does not necessarily result in a satisfactory reproduction of the cross-section data\footnote{This is partly related to the low cut-off energy $\Lambda$ which is due to the limited applicable energy domain of the low-energy model.}. 
We will therefore fit a LHS-\textit{inspired} $t$-dependence of the Regge pole residues to the high-energy data and subsequently evaluate the RHS of the FESR in Eq.~\eqref{eq:FESR_main}. The latter is then compared to the LHS of the FESR.

\begin{figure}[tbh]
\centering
\includegraphics[width=2.5cm]{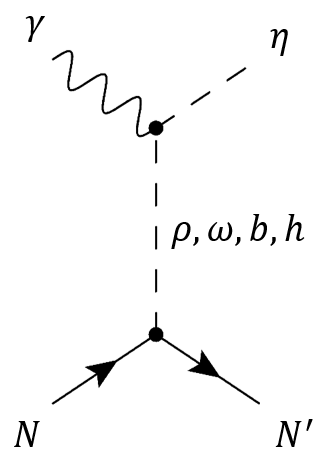}
\caption{Single-meson $t$-channel exchange diagram.\label{fig:t_channel_exchange_single_particle}}
\end{figure}

To obtain a better intuition about the Regge exchange parameters  entering the scalar amplitudes $A_i$ we compute those using a particle exchange instead of a Reggeon exchange model
({\it c.f.}  Fig.~\ref{fig:t_channel_exchange_single_particle}). 
For example using $R^\rho =  1/(t-m_\rho^2)$
 for the $\rho$ meson exchange, we obtain the following contributions to $A^{\rho}_i$~\cite{Mathieu:2015eia}
\begin{align}
A^{\rho}_1 = g^{\rho}_1 t R^{\rho} \,,~ A^{\prime \rho}_2 = 0 \,,~ A^{\rho}_3 = 0 \,,~ A^{\rho}_4 = g^{\rho}_4 R^{\rho}\,.
\end{align}
In the $s$-channel, $g_1$ ($g_4$) corresponds to a nucleon-helicity flip (non-flip). For $b$ meson exchange 
\begin{align}
A^{b}_1 &= 0  \,,~ A^{\prime b}_2 = t g^b_2 R^b \,,~A^{b}_3 = 0 \,,~A^{b}_4 = 0\,.
\end{align}
These indicate the $t$-factors that are necessary for angular-momentum conservation and factorizable $t$-channel exchanges. It should be stressed however, that Regge residue factorization is a stronger constraint than factorization of on-shell couplings since the former imposes a relation among the residues for all $t$. 

Among others, the above effective parameters will later be fitted to the available high-energy data. Below we derive estimates for the coupling constants in order to constrain the fit to realistic values.
It will be useful for comparison to relate the couplings $g_1$ and $g_4$ to the standard electromagnetic tensor $g_t$ and vector $g_v$ coupling constants, $\lambda_{V\eta\gamma}$~\cite{Chiang:2001as,Benmerrouche:1994uc} 
\begin{align}\label{eq:relations_couplings_trad_vm}
g_1^V = \frac{e \lambda_{V\eta\gamma}}{\mu} \frac{g_t^V}{2M_N} \,, \quad g_4^V =- \frac{e \lambda_{V\eta\gamma}}{\mu} g_v^V \,.
\end{align}
As an initial estimate we take the coupling constant for the $\rho$ and $\omega$ exchange from the $\eta$-MAID model (see Table~2 in Ref.~\cite{Chiang:2001as})
\begin{align}
\lambda_{\omega \eta \gamma} &= 0.29 \,,&g_{v}^\omega &= 16.0 \,, &g_t^\omega &= 0 \,,\\
\lambda_{\rho \eta \gamma} &= 0.81 \,,&g_{v}^\rho &= 2.4 \,, &g_t^\rho &= 14.64 \,.
\end{align}
Note that $\lambda_{\rho \eta \gamma} \approx 3 \lambda_{\omega \eta \gamma}$ as expected from SU(3) flavor symmetry. 
These couplings are related to the $g^V_1$ and $g^V_4$ according to Eq.~\eqref{eq:relations_couplings_trad_vm} yielding
\begin{align}
&g^\omega_1 = 0 \,,  &g^\omega_4 = -2.57 \mbox{ GeV}^{-1} \,,\\ 
&g^\rho_1 = 3.49 \mbox{ GeV}^{-2} \,, &g^\rho_4 = -1.07 \mbox{ GeV}^{-1} \,.
\end{align}
These estimates show that $\omega$ is expected to be dominantly helicity non-flip, while $\rho$ is dominantly helicity flip. This is consistent with fits to the high-energy data from the relative helicity-flip and non-flip $F/D$ ratios in combination with SU(3) flavor symmetry  (see for example Table AA.4c in Ref.~\cite{Irving:1977ea}). For the $b$ and $h$ exchange little is known about their couplings~\cite{reviewofparticlephysics} and 
 the $\eta$-MAID model does not include these exchanges. We obtain a first estimate based on the predictions from the low-energy side of the FESR. 
In Appendix~\ref{sec:b_coupling}, we obtain $g_2^b = 3.80\mbox{ GeV}^{-2}$  for the $b$-coupling based on SU(3) flavor symmetry and vector meson dominance.

\begin{figure}[thb]
\centering
\includegraphics[width=0.3\textwidth]{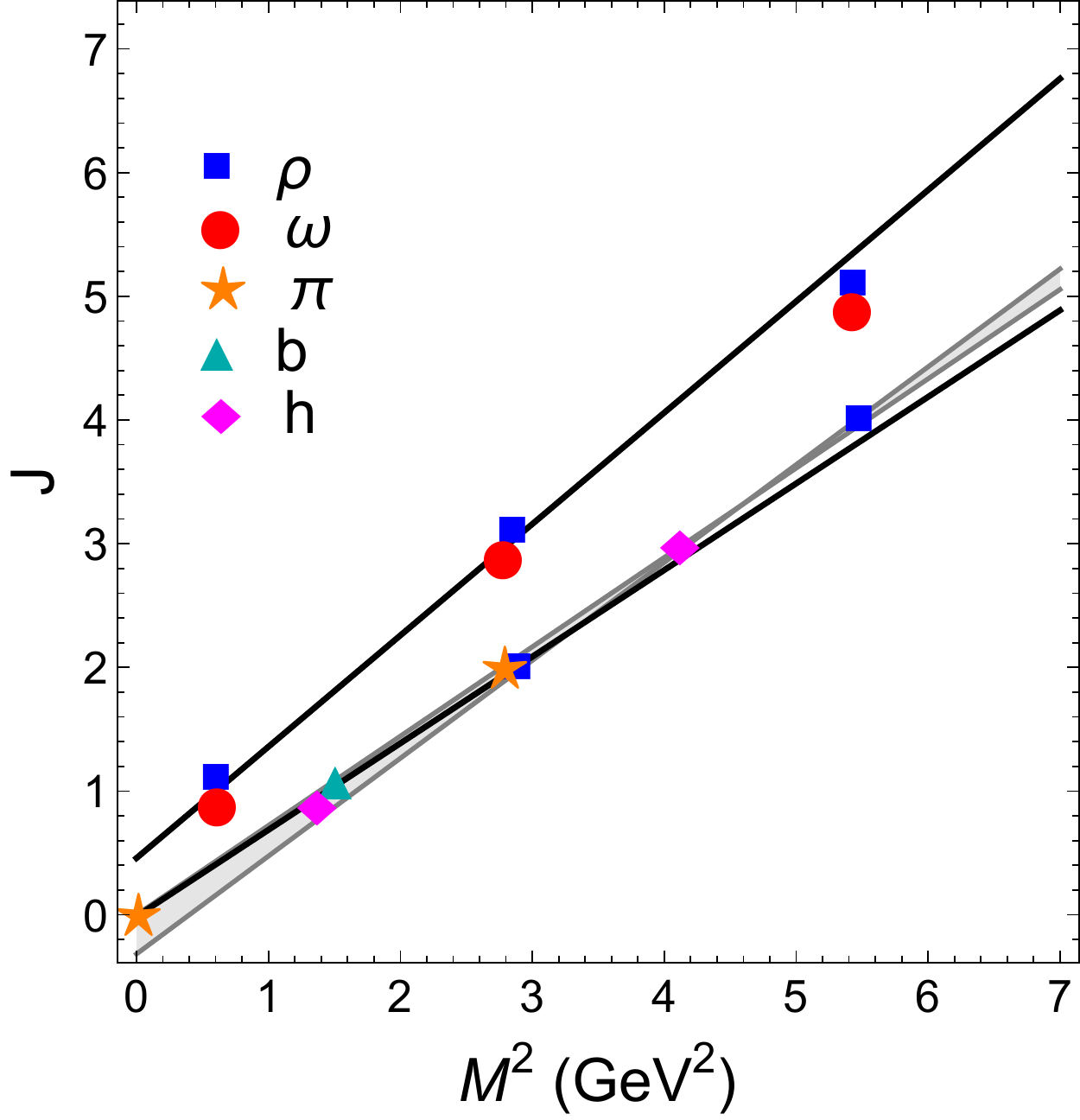}
\caption{Chew-Frautchi plot including the $\pi$, $b$, $h$ and $\rho$ excitations and quark model states $2^{--}$ and $4^{--}$. The black lines show the trajectories $\alpha(t) = 0.7(t-m_\pi^2)$ and $\alpha(t) = 1+0.9(t-m_\rho^2)$.
\label{fig:trajectory}}
\end{figure}

For the Regge exchange the trajectories $\alpha^\rho$ and $\alpha^\omega$ are fixed by considering the resonance  spectrum. This fixes the $s$-dependence of the Regge-pole contributions. For $V=\rho,\omega$, we assume weak degeneracy $\alpha^V(t) = 1+\alpha^{\prime V} (t - m_\rho^2)$ where $\alpha^{\prime V} = 0.9\mbox{ GeV}^{-2}$. 
Note that $\alpha^V(t_0) = 0$ for $t_0 \approx -0.5 \textrm{ GeV}^2$. For the axial vectors $A = b,h$ we assume weak degeneracy with the pion trajectory $\alpha^A(t) = \alpha^{\prime A} (t - m_\pi^2)$ with $\alpha^{\prime A} = 0.7\mbox{ GeV}^{-2}$ (see Fig.~\ref{fig:trajectory}).
Within the  Regge-pole model a number of constraints can be derived for the $t$-dependence of Regge residues  $\beta_i^\sigma(t)$ by comparing with the LHS of the FESR. 
 The two sides are compared in Fig.~\ref{fig:lhs_vs_rhs_2} and below  we summarize the main findings.
 
\begin{figure}[htb]
\centering
\includegraphics[width=0.3\textwidth]{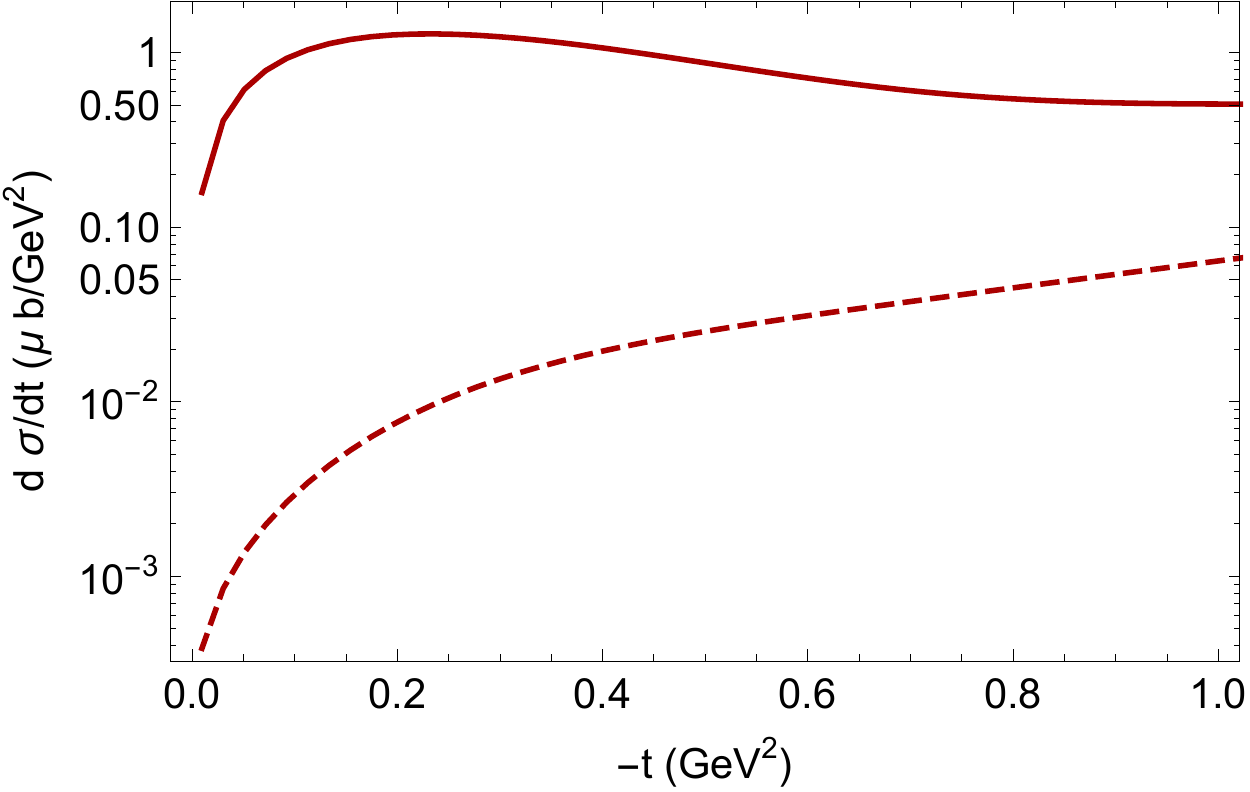}
\caption{The $\eta$-MAID differential cross section at $W = 2.0$ GeV. The solid line is the full model, and the dashed line shows the $A_3$ contribution.\label{fig:effect_of_A3_EMAID}}
\end{figure}

\begin{itemize}

\item The vector and axial-vector Regge amplitudes in Eq.~\eqref{eq:regge_R} have poles at odd integer values of $\alpha$. The poles generated by the $\sin \pi \alpha$ denominator at even integer $\alpha$ are removed by the signature factor $1-e^{-i\pi\alpha}$. Poles located at negative integer $\alpha$ are unphysical and should be canceled by residue zeros. Such poles can be removed by taking $\beta \propto 1/\Gamma(\alpha+1)$ but this parametrization is not unique, {\it e.g.} one can write $\beta(t) \propto (\alpha+1)(\alpha+2)(\alpha+3)...$ which in combination with the signature factor, forces the amplitude to be finite (zero) at negative odd (even) integer $\alpha$.

\item A single Regge pole with $\alpha = 0$, physically  corresponds to a spin-$0$,  $t$-channel exchange. For the $\rho$ and $\omega$ trajectories, this corresponds to $t \approx -0.5 \textrm{ GeV}^2$. At $\alpha = 0$, the signature factor removes the wrong-signature pole generated by $\sin \pi \alpha$, but the amplitude remains finite. Since a spin-$0$ exchange cannot flip the nucleon helicity, the Regge residues in the $t$-channel spin-flip amplitudes are expected to vanish at $t \approx -0.5\textrm{ GeV}^2$. These are referred to as the nonsense wrong signature zeros (NWSZ). Similar zeros are expected, for example in $\pi^0$ photoproduction amplitudes~\cite{vincent_fesr_pion}. 
Assuming factorization, the hadronic vertex in neutral meson photoproduction reactions can be related to $\pi N$ scattering residues. A zero has also been observed in the $t$-channel isovector helicity-flip amplitude $B^{(-)}$ in a recent FESR analysis of low-energy $\pi N$ scattering models~\cite{Mathieu:2015gxa}.

\item The definite parity, singularity free $t$-channel helicity-flip amplitudes can be written in terms of the invariant amplitudes as~\cite{Mathieu:2015eia}
\begin{align}
F_3 &= 2M_N A_1 - t A_4 \,, \label{eq:F_3_t_ch} \\
F_4 &= A_3 \,.
\end{align}
The FESR for these amplitudes are depicted in Fig.~\ref{fig:t_ch_hel_flip_FESR}.
For $\omega$ exchange the nonsense wrong signature zeros are clearly present in 
$\textrm{LHS}(A_4^s)$ and $\textrm{LHS}(A_3^s)$ and possibly in $\textrm{LHS}(A_1^s)$. However for the $\rho$, only the $\textrm{LHS}(A_4^v)$ has the zero, while $\textrm{LHS}(A_1^v)$ and $\textrm{LHS}(A_3^v)$ are finite near $t \approx -0.5 \textrm{ GeV}^2$. The absence of the NWSZ for the $\rho$ exchange suggests the importance of non-factorizable corrections in this amplitude.

 \item The presence or absence of NWSZ distinguish  $\pi^0$ from $\eta$ photoproduction. In $\pi^0$ photoproduction, there is a dip in the cross section near $t \sim -0.5\mbox{ GeV}^2$ because of the zero for the exchange $\omega$, which is dominant there (see Eq.~\eqref{eq:gault_relation}), while for $\eta$, the $\rho$ is dominant which does not have this dip. 
 
\item  One can force the NWSZ by taking $\beta(t) \propto \alpha(t)$ in the corresponding $t$-channel helicity flip amplitudes. This procedure is referred to as the nonsense mechanism~\cite{collins,Mathieu:2015gxa}. 
Since $\omega$ is dominantly $s$-channel helicity non-flip, {\it i.e.} $A_1^\omega \sim 0$,  one can approximate $F_3^\omega = -t A_4^\omega$ and so we take $\beta_4^\omega \sim \alpha$ for simplicity. Since there is no NWSZ in $A_1^\rho$,  there is no need to impose such a relation between $\beta$ and $\alpha$ for  $F_3^\rho$. However, since a zero is observed in $\textrm{LHS}(A_4^\rho)$, we do impose $\beta_4^\rho \sim \alpha$. 

\begin{figure}[tbh]
\centering
\includegraphics[height=35mm]{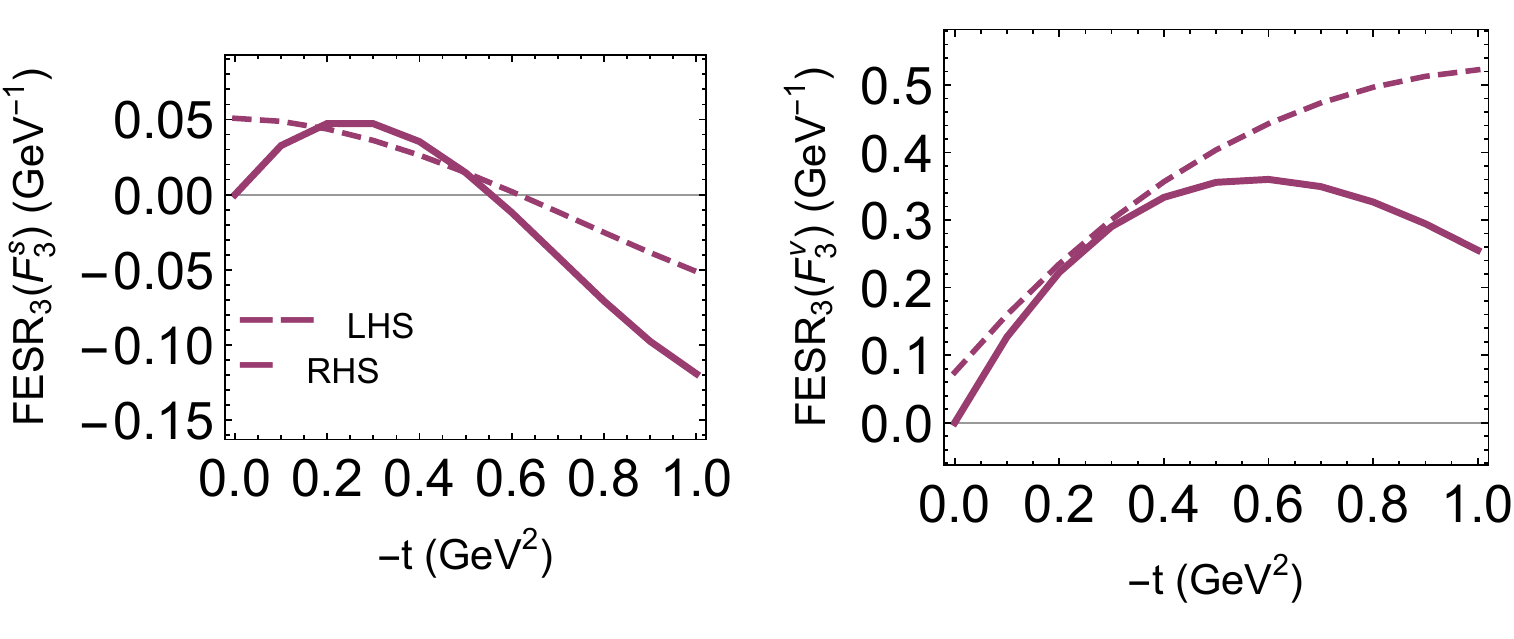}
\caption{LHS and RHS of the FESR for the two isospin components of the definite parity, singularity free $t$-channel nucleon-helicity flip amplitude $F_3$ (see Eq.~\eqref{eq:F_3_t_ch}) \label{fig:t_ch_hel_flip_FESR}.}
\end{figure}

\item Since the $h$ and $b$ exchanges have quantum numbers corresponding to $t$-channel nucleon-helicity non-flip only, no NWSZ are expected in their residues. The $\alpha^b = 0$ occurs at $t = 0.018 \textrm{ GeV}^2$ and indeed neither of the $\textrm{LHS}(A_2^{s,v})$ suggest the presence of a zero at this $t$.

\item As discussed above, there are no known Reggeons that would contribute to $A_3$. The $A_3$ corresponds to quantum numbers of unnatural exchanges. However, as seen from the LHS of the FESR, this contribution is non-negligible. Figure~\ref{fig:effect_of_A3_EMAID} shows the contribution of the $A_3$ to the cross section evaluated using the $\eta$-MAID 2001 model. At small $-t$, its contribution is small 
 increasing towards larger values of $-t$. In this section we discuss the high energy parametrization where a  `conservative model' is presented. The model consists solely of known exchanges and for which $A_3^{s,v} \equiv 0$. In the next section, we elaborate an `exploratory model' where we study the possibility of including Regge trajectories for mesons which, albeit predicted by lattice QCD and quark models~\cite{Godfrey:1985xj,Dudek:2013yja}, have not been observed yet. 
\end{itemize}

According to the arguments presented above we use the following parametrization for vector contributions $V = \rho,\omega$ (using the notation of Eq.~\eqref{eq:regge_R})
\begin{align}
\beta^V_1(t) &= g^V_1 t \frac{-\pi \alpha^{\prime V}}{2 }\frac{1}{\Gamma(\alpha^V(t) + 1)} \label{eq:beta_1_V} \,,\\
\beta^V_4(t) &= g^V_4 \frac{-\pi \alpha^{\prime V}}{2 }\frac{1}{\Gamma(\alpha^V(t))}  \label{eq:beta_4_V} \,
\end{align}
while for the axial vectors $A = b,h$ we use
\begin{align}
\beta^{\prime A}_2(t) &= g^A_2 t \frac{-\pi \alpha^{\prime A}}{2 }\frac{1}{\Gamma(\alpha^A(t)+1)} \label{eq:b_gamma_plus_one} \,,
\end{align}
where the prime in $\beta_2'$ denotes the fact that this is the $A_2'$ residue. This also explains the factor of $t$. The factor $-\pi \alpha^{\prime}/2$ ensures the correct on-shell couplings.
The functions $1/\Gamma(\alpha+1)$ and $1/\Gamma(\alpha)$ are both equal to $1$ at the pole $\alpha = 1$, yet they result in a different strength in the physical region. 
The scale parameters $r_i$ ({\it cf.} Eq.~\eqref{eq:regge_R})  are found to efficiently compensate for the increased strength brought in by $1/\Gamma(\alpha + 1)$ and allows one to hold on to the on-shell couplings calculated earlier. The $r_i$ parameters  affect the slope in $t$ of the amplitudes by introducing an exponential damping $\exp \left[ (\alpha(t) - 1) \ln r \right]$ at large $-t$ since we take $r \geq 1 \mbox{ GeV}^{-1}$ and $\alpha(t) < 1$ in the physical region.


Each exchange $e$ is assigned its own scale parameter in Eq.~\eqref{eq:regge_R}, which will be denoted by $r_i^e$. We introduce the parameter reductions $r_1^\rho= r_1^\omega$, $r_4^\rho= r_4^\omega$ and $r_2^b= r_2^h$, and, therefore, we drop the superscript. In order to further reduce the number of free parameters, we assume weak degeneracy and relate the coupling of the $h$-meson to the coupling of the $b$-meson by $g_2^h=2 g_2^b/3$, according to the LHS predictions (see Section~\ref{sec:lhs}).

\begin{figure}[tbh]
\centering
\includegraphics[width=0.3\textwidth]{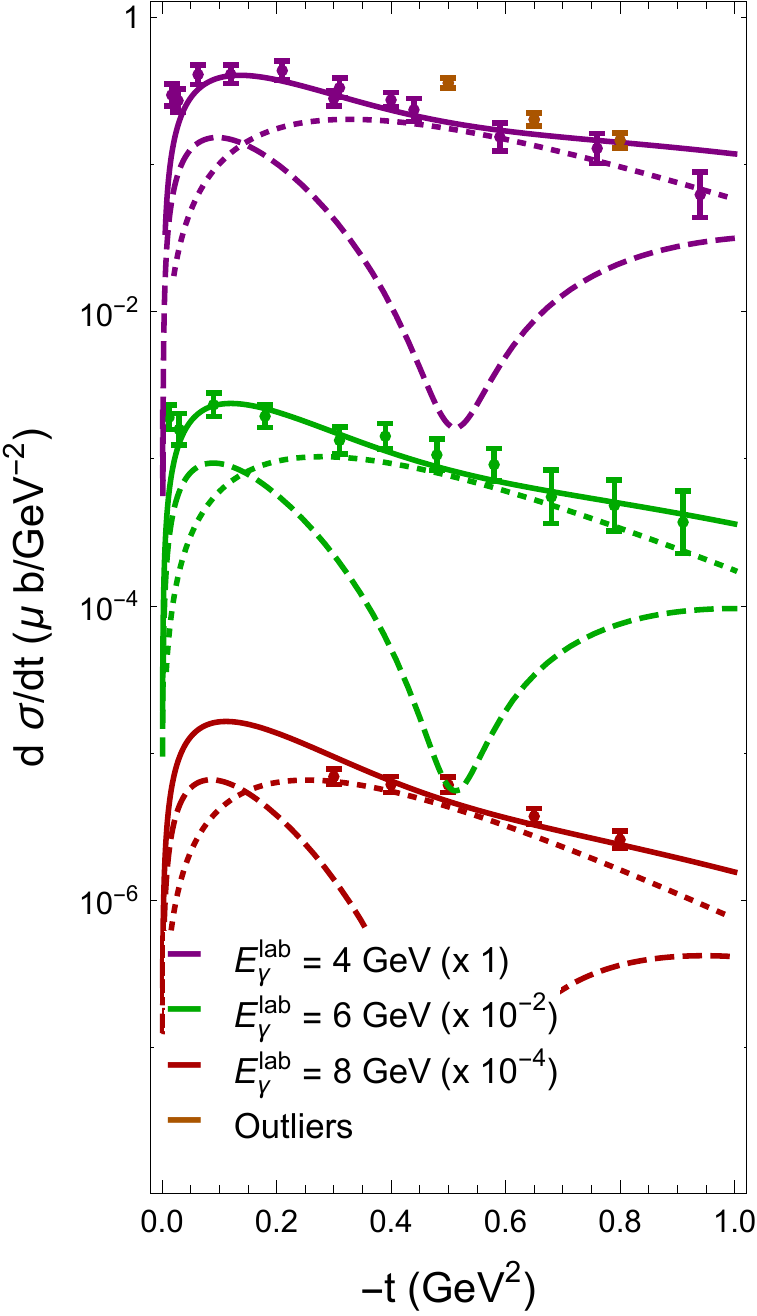}
\caption{High energy $\gamma p \rightarrow \eta p$ data compared to the fitted Regge model. The dotted (dashed) line shows the isovector (isoscalar) contribution. The solid line represents the full Regge model. Data are from Refs~\cite{Braunschweig:1970jb,Dewire:1972kk}. The three data points in brown from Dewire~\textit{et al.}~\cite{Dewire:1972kk} at $E_\gamma^{\textrm{lab}} = 4$ GeV were excluded from the fit due to a systematic inconsistency.\label{fig:cross_section}}
\end{figure}

\begin{figure}[htb]
\centering
\includegraphics[width=0.4\textwidth]{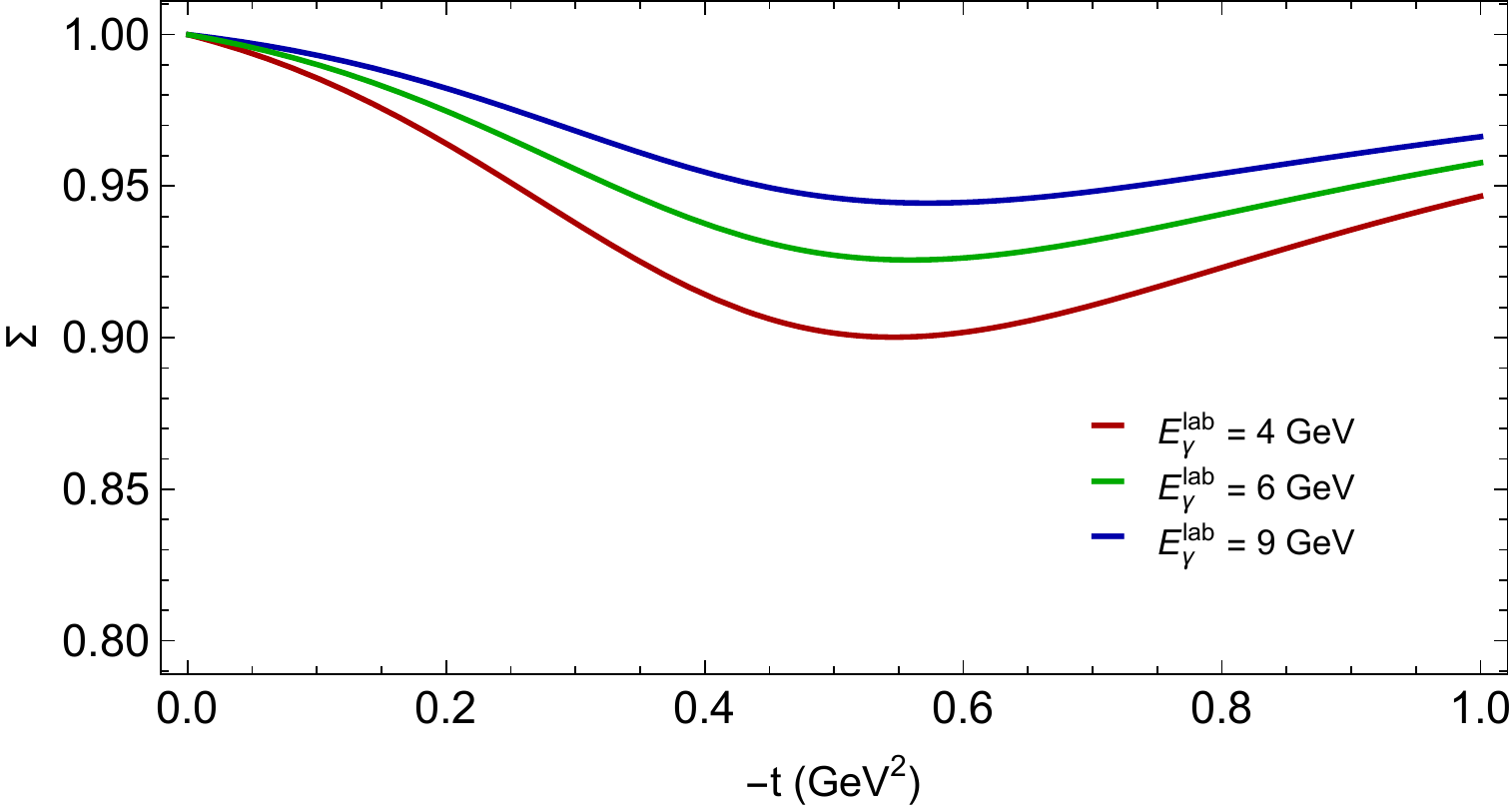}
\caption{Predictions for the beam asymmetry at $E_\gamma^{\textrm{lab}} = 4,6,9$  GeV for the conservative model.\label{fig:gluexSigma}}
\end{figure}

So far we have discussed a high-energy model which incorporates the features observed in the low-energy predictions of the residues. We have provided realistic estimates for the coupling constants of the leading $t$-channel exchanges in order to set the scale of the individual contributions. These estimates are necessary when, in the next step, we fit the high-energy model to the available cross-section data by varying coupling constants and scale parameters. We use data for $E_\gamma^{\textrm{lab}} \geq 4 \textrm{ GeV}$  from~\cite{Braunschweig:1970jb,Dewire:1972kk} (for details see Fig.~\ref{fig:cross_section}). Since the number of high-energy data points is rather limited (31 cross section measurements at 3 different beam energies)  we constrain the couplings within a predefined range centered around the estimates given earlier, in order to avoid overfitting data. The $s$ dependence of the model is fixed by the Regge trajectories so only the $t$ dependence and strength of the contributions is allowed to vary.
Our model involves eight free parameters: five coupling constants $g_1^\rho$, $g_1^\omega$, $g_4^\rho$, $g_4^\omega$ and $g_2^b$ and three scale parameters $r_1$, $r_2$ and $r_4$. The coupling constants are constrained within $30 \%$ around the values estimated above. The exception is $g_1^\omega$, which we constrain to be in  the range $0 \leq g_1^\omega \leq 0.2\mbox{ GeV}^{-2}$. The scale parameters may assume all values greater than or equal to one. The optimal parameters are given Table~\ref{tab:regge_coefficients_fit} which correspond to $\chi^2/\mbox{d.o.f.}=3.04$. The largest contributions to the $\chi^2$ are related to the data at very forward scattering angles $-t < 0.1 \mbox{ GeV}^2$. It should be noted that the cross section fit does not force hard constraints on $g_2^b$. The resulting model is compared to the data in Fig.~\ref{fig:cross_section} and beam-asymmetry predictions are presented in Fig.~\ref{fig:gluexSigma}. 

\begin{table}[htb]
\caption{Parameter values of the high-energy model obtained from a constrained $\chi^2$ minimization.\label{tab:regge_coefficients_fit}}
\begin{tabular}{c|c|c}
 \textbf{Parameter} & \textbf{Fit} &\textbf{Initial estimates}\\
 \hline\hline
 $g^\rho_1$ &  $\phantom{+}3.434 \pm 0.083\mbox{ GeV}^{-2}$ & $\phantom{+}3.49\mbox{ GeV}^{-2}$\\
 $g^\rho_4$ & $-1.397 \pm 0.085\mbox{ GeV}^{-1}$ & $-1.07\mbox{ GeV}^{-1}$\\
  $g^\omega_1$ & $\phantom{+}0.116 \pm 0.074\mbox{ GeV}^{-2}$ & $0$ \\
 $g^\omega_4$ & $-3.346 \pm 0.087\mbox{ GeV}^{-1}$ & $-2.57\mbox{ GeV}^{-1}$\\
 $g^b_2$ & $\phantom{+}4.946 \pm 1.491\mbox{ GeV}^{-2}$ & $\phantom{+}3.80\mbox{ GeV}^{-2}$\\
 $r_1$ & $\phantom{+}3.001 \pm 0.087\mbox{ GeV}^{-1}$ & -\\
 $r_4$ & $\phantom{+}1.974 \pm 0.101\mbox{ GeV}^{-1}$ & - \\
 $r_2$ & $\phantom{+}6.204 \pm 2.484\mbox{ GeV}^{-1}$ & -
\end{tabular}
\end{table}

The fit fixes the residues of the high-energy model. Plugging the results in the RHS of Eq.~\eqref{eq:FESR_main}, we obtain the high-energy prediction of the sum rules. The latter can be compared to the LHS of the sum rules, originating from the low-energy model. The RHS of the $A_4^{s,v}$ amplitudes show the same shape for the residues as predicted by the LHS, but is not able to reproduce the sign of one of the $\beta_4^{s,v}$ couplings. The cross section on a proton target can be decomposed at leading order in $s$ as follows:
\begin{align}
\frac{d \sigma}{dt} = \frac{1}{32 \pi} \left( \abs{A_1^\omega + A_1^\rho}^2 - t \abs{A_4^\omega + A_4^\rho}^2 + \abs{A_2^{\prime b} + A_2^{\prime h}}^2 \right) \,.
\end{align}
Because of the assumed degeneracy of the  $\rho$ and $\omega$ trajectories and residues $\beta_4^{\rho, \omega}$ we cannot isolate their individual contributions. Since only proton target ${d\sigma}/{d t}$ information is available, our fit is only  sensitive to $\abs{A_4}^2 \sim  \abs{g^\omega_4 + g^\rho_4}^2$. 
The LHS of the FESR suggest a destructively interfering isoscalar and isovector contribution to the $A_4$. In our high-energy model, $A_4^s$ and $A_4^v$ require the same sign in order to properly reproduce the forward bump around $t = -0.1 \mbox{ GeV}^2$. It is not clear from the available high-energy data which isospin component should have an opposite sign compared to its LHS prediction.

We cross-check this sign inconsistency between the LHS and RHS predictions with other models. The comparison between the high-energy proton amplitudes and a number of low-energy models is depicted in Fig.~\ref{fig:regge_comparison_to_models}. It is clear from these figures that the $A_4$ amplitude is ill-constrained among the low-energy models, making it unclear whether the $A_4$ inconsistency is due to the choice of the $\eta$-MAID model, or rather to a shortcoming of the high-energy parametrization. The best agreement at low $-t$ and $W \leq 2\mbox{ GeV}$ is obtained with the Bonn-Gatchina model, which Reggeizes the $t$-channel contributions. All resonances contribute to the $A_4$ amplitude (see Eq.~\eqref{eq:A4_ifo_Fi}), making it highly sensitive to the model assumptions. Finally, it should also be noted that, while the LHS does not match the RHS, the couplings $g_4^\rho$ and $g_4^\omega$ do have the same sign as the $t$-channel contributions in the low-energy model. 
One might argue that the missing strength in the forward direction is related to imposing factorization of the $\omega$ contribution in the $A_1^\omega$. However, we find that when the constraint of a vanishing $A_1^s$ at $t = 0$ is removed, one is unable to reproduce the forward bump at $-t \approx 0.1 \mbox{ GeV}^2$ when $A_4^s$ and $A_4^v$ contribute with opposite sign.

The $A^{\prime s}_2$ and $A^{\prime v}_2$ are found to be small and represent a negligible contribution to the cross section. However, the unnatural contributions cannot be neglected since they can be clearly identified in the beam-asymmetry ($\Sigma$) in accordance with Stichel's theorem~\cite{Bajpai:1971zf}. At leading order, one obtains
\begin{align}
\Sigma &= \frac{ \left( \abs{A_1}^2 - t \abs{A_4}^2\right) - \left(\abs{A_2'}^2 - t \abs{A_3}^2\right) }{\left( \abs{A_1}^2 - t \abs{A_4}^2 \right) + \left(\abs{A_2'}^2 - t \abs{A_3}^2\right)} \label{eq:Sigma_schematically}\,.
\end{align}
Hence, for a dominating natural exchange ($A_1$ and $A_4$), $\Sigma = +1$ is expected, while purely unnatural exchange ($A_2'$ and $A_3$) corresponds to $\Sigma = -1$. According to factorization, all amplitudes must vanish as $t \rightarrow 0$. Bearing in mind the $t$-factors both explicitly and implicitly written in Eq.~\eqref{eq:Sigma_schematically}, the expected behavior in both cases at small $t$ is
\begin{align}
\Sigma &\underset{t\rightarrow 0}{\sim} \frac{\abs{A_1}^2-\abs{A_2'}^2}{\abs{A_1}^2+\abs{A_2'}^2} \label{eq:Sigma_for_ang_mom_cons}\qquad \textrm{(ang.\ mom.\ conservation)} \\
\Sigma &\underset{t\rightarrow 0}{\sim} \frac{\abs{A_4}^2-\abs{A_3}^2}{\abs{A_4}^2+\abs{A_3}^2} \label{eq:Sigma_for_factorization}\qquad \textrm{(factorization)} 
\end{align}
We show our predictions for the beam asymmetry at $E_\gamma^{\textrm{lab}} = 9$ GeV in Fig.~\ref{fig:gluexSigma}. Some important remarks can be made here. Since the current model is dominated by natural exchange, the result is close to $\Sigma = +1$. At $t \approx -0.5 \textrm{ GeV}^2$, a dip is observed, which is generated by the vanishing $A_4$ contribution from natural exchange. Assuming factorization and $A_3 \equiv 0$, only $\Sigma = +1$ is possible at $t = 0$. Any experimentally observed deviation suggests either an $A_3$ contribution or a violation of factorization. The experimental signature of both possibilities will be demonstrated in the next section.

Our only reference of the relative isospin contributions in the high-energy data is the strength of the LHS of the FESR. The upcoming GlueX results on photon asymmetries in both pion and eta photoproduction would represent an invaluable source of information in this respect. 
For example, in a combined analysis one may be able to learn about the $h$ contribution. 
The relative size of $\Sigma(\gamma p \rightarrow \eta p)$ and $\Sigma(\gamma p \rightarrow \pi^0 p)$ at the same kinematics is related to the relative strength of the unnatural isoscalar and isovector exchanges in a Regge-pole model~\cite{Irving:1977ea}. Considering Eq.~\eqref{eq:gault_relation}, it can easily be seen that the isoscalar contributions are suppressed by a factor of $9$ compared to the isovector contributions in $\eta$  photoproduction, relative to $\pi^0$ photoproduction. By comparing the beam asymmetry in both channels, one can extract the relative strength of the contributions.

\begin{figure}[htb]
\centering
\includegraphics[width=7cm]{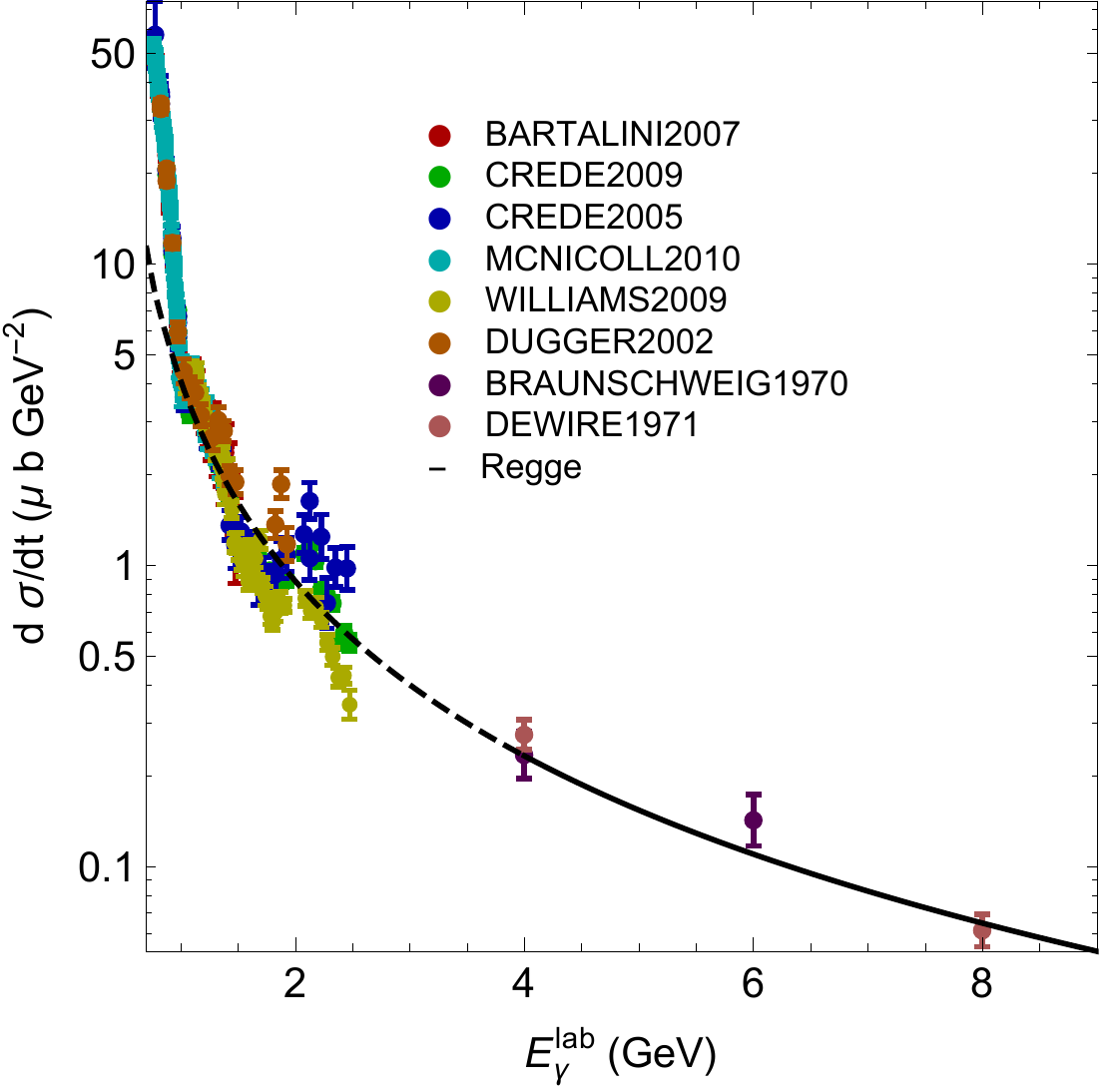}
\caption{Extrapolation (dashed line) of the high-energy model, which is valid from $E_\gamma^\textrm{lab} = 4.0$ GeV (solid line). The model is evaluated at $t = -0.4 \textrm{ GeV}^2$ and data is collected for $0.35 \leq -t \leq 0.45 \textrm{ GeV}^2$. Data are from Refs.~\cite{Bartalini:2007fg,Crede:2009zzb,Crede:2003ax,McNicoll:2010qk,Williams:2009yj,Dugger:2002ft,Braunschweig:1970jb,Dewire:1972kk}.\label{fig:lowEplot}}
\end{figure}

For completeness, we compare the high-energy model (valid for $E_\gamma^\textrm{lab} \geq 4 \textrm{ GeV}$) to the available low-energy data in Fig.~\ref{fig:lowEplot}, where the model is extrapolated outside its scope of application. The Regge model reproduces the low-energy data on average (except close to threshold), illustrating the fact that also the real parts of the high-energy amplitudes are consistent with low-energy data.

\section{An exploratory model}\label{sec:exploratory}
\begin{figure}[thb]
\centering
\includegraphics[width=0.3\textwidth]{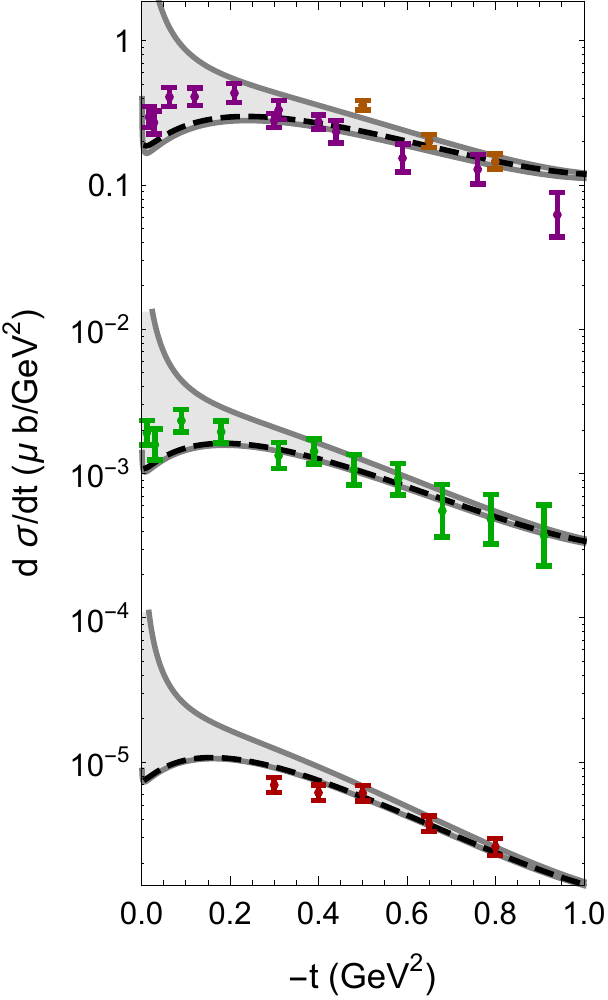}
\caption{Predictions for the cross-section within the extended model for $k=2,3$. The band corresponds to various $\rho_2$ and $\omega_2$ trajectories. The data are scaled as in Fig.~\ref{fig:cross_section}.\label{fig:diffcs_from_lhs}}
\end{figure}

Since the LHS of the FESR suggest a non-negligible $A_3$ component, we consider an alternative description of the high-energy amplitudes that is further constrained by the low-energy prediction for the sum rules. By considering the residues and scale factors
\begin{align}\label{eq:low_energy_beta}
\beta^\sigma_{i}(t) r_i^{\alpha(t)-1}= \textrm{LHS}_k(A_i^\sigma) \frac{\alpha(t)+k}{\Lambda^{\alpha(t)-1}} \,,
\end{align}
one can construct a Regge-pole model directly from the low-energy model with minimal assumptions. In order to compute the residues, $\beta^{s,v}_3$ from Eq.~\eqref{eq:low_energy_beta}, one also needs a model for the 
 corresponding Regge trajectories. In absence of experimental information we base our estimate of the trajectory functions on the quark model predictions. In both the isoscalar and isovector case, a relativized quark model~\cite{Godfrey:1985xj} predicts two states with masses\footnote{The states reported in Ref.~\cite{Anisovich:2002su} have masses $m_{\rho_2} = 1.94 \mbox{ GeV}$, $m_{\rho_4} = 2.23 \mbox{ GeV}$ and those in Ref.~\cite{Anisovich:2011sva} have masses $m_{\omega_2} = 1.97\mbox{ GeV}$ and $m_{\omega_4} = 2.25\mbox{ GeV}$. Hence, these states suggest a much steeper trajectory with an intercept $\alpha = 0$ at higher $t$.}
$m_{2^{--}} = 1.7 \textrm{ GeV}$ and $m_{4^{--}} = 2.34 \textrm{ GeV}$, which, assuming a linear trajectory, leads to $\alpha(t) = -0.235 + 0.774 t$. The states are depicted in the Chew-Frautschi plot in Fig.~\ref{fig:trajectory} where we notice a compatibility with the $b$ and the $\pi$ trajectories. 
The high-energy amplitude is sensitive to variations in the trajectory slope and intercept. It should be noted, that for $t$ in the range $0 \leq -t \leq 1 \textrm{ GeV}^2$ such that 
 $\alpha(t) = 0$, the amplitude has an unphysical pole which needs to be canceled by residue zeros. For the isoscalar and isovector part of the LHS, a zero is found at $t \approx -0.7 \textrm{ GeV}^2$ and $t \approx 0.3  \mbox{ GeV}^2$ respectively. These zeros impose a relation between the slope and intercept of the trajectories if they are assumed to be related to the $\alpha = 0$ point. In the case of the isoscalar amplitude, the restriction $\alpha(t = -0.7 \mbox{ GeV}^2) = 0$ has poor correspondence to the quark-model states. For the isovector part on the other hand, the constraint $\alpha(t = 0.3  \mbox{ GeV}^2) = 0$ is in good agreement with the quark model, which predicts $\alpha(t = 0.304 \mbox{ GeV}^2) = 0$. To study the trajectory dependence of the high-energy model, we extract the residues $\beta_3^{s,v} $ in Eq.~\eqref{eq:low_energy_beta} using a range of trajectories. We vary the location of the pole $\alpha=0$ within the range $0 \leq t \leq 0.4 \mbox{ GeV}^2$ and determine the trajectory slope and intercept by a least-square fit to the quark-model states. The range of trajectories is shown in Fig.~\ref{fig:trajectory}. The effect on the cross section is illustrated in Fig.~\ref{fig:diffcs_from_lhs}. The main experimental sensitivity is at small $-t$, where a pole close to the physical region overestimates the data (where it is not canceled by a residue zero). For a distant pole, the effect of the $A_3$ contributions is negligible. It should be noted that the $\alpha = 0$ point corresponds to an exotic $0^{--}$ state. The increased cross section at low $-t$ is a manifestation of this state. While interesting experimentally, we do not expect such a signature to be seen in high-energy experiments.

\begin{figure}[thb]
\centering
\includegraphics[width=0.3\textwidth]{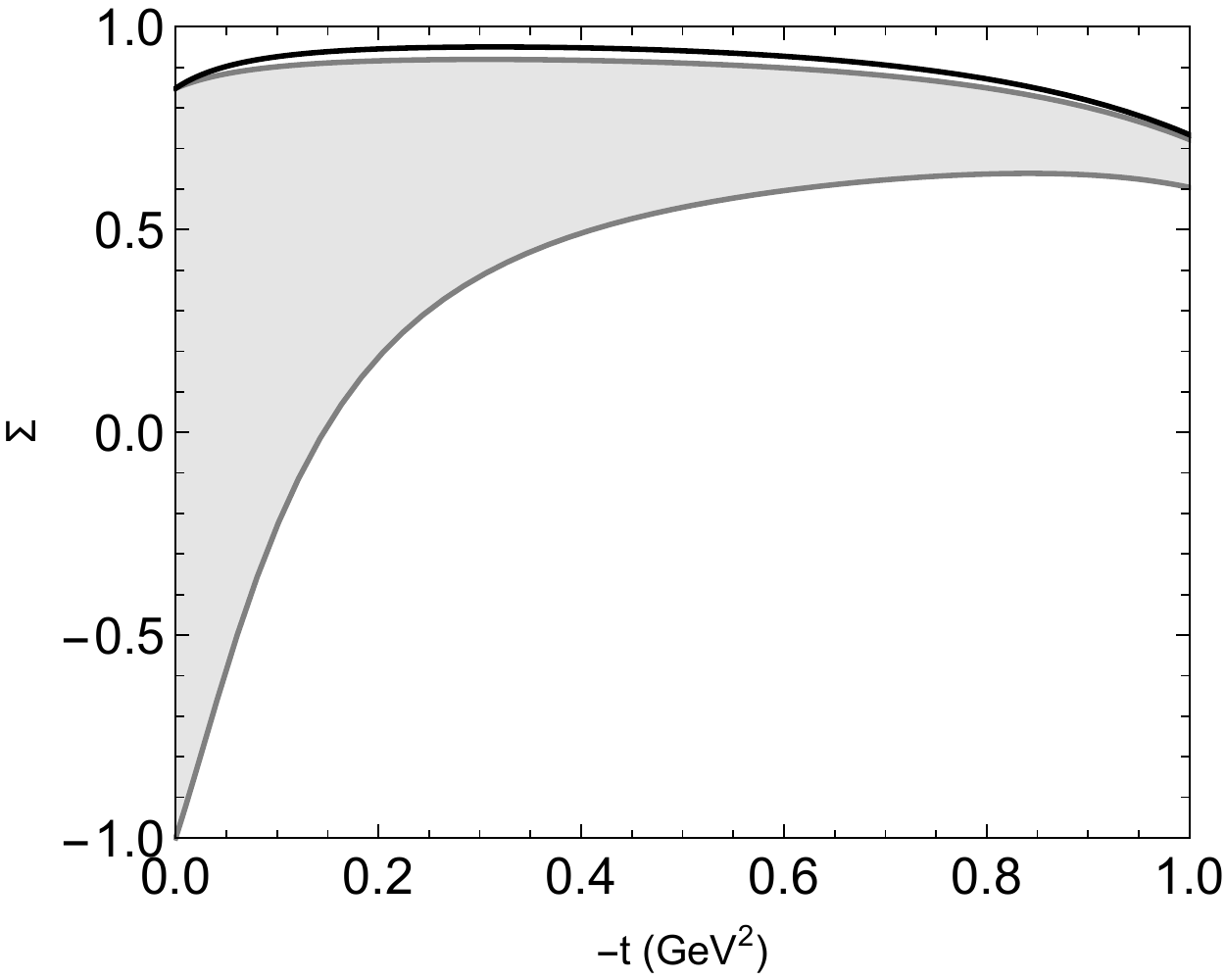}
\caption{Predictions for the beam asymmetry at $E_\gamma^{\textrm{lab}}=9 \textrm{ GeV}$ for $k=2,3$ without (black solid curve) and with (gray band) the $A_3$ contributions for various $\rho_2$ and $\omega_2$ trajectories.\label{fig:sigma_from_lhs}}
\end{figure}

Using the same procedure, we study the effect on the beam asymmetry induced by the uncertainty of the trajectory in Fig.~\ref{fig:sigma_from_lhs}. In the conservative model, which incorporates factorization explicitly, $\Sigma = +1$ is obtained at $t = 0$, in agreement with Eq.~\eqref{eq:Sigma_for_factorization}. The signature of factorization is now clear in Fig.~\ref{fig:sigma_from_lhs} where $\Sigma$ is slightly smaller than $+1$.  Switching on the $A_3$ contribution generates a strong dip at forward angles. The further away from the physical region the $\alpha = 0$ is located, the weaker is the contribution from $A_3$. In the latter case, the beam asymmetry is closer to  $+1$.



\section{Conclusions and outlook}\label{sec:summary}
We have analyzed $\gamma N \rightarrow \eta N$ using the framework of finite-energy sum rules. 
Using these sum rules, one is able to obtain the $t$-dependence of the high-energy Regge residues using low-energy models. 
We found zeros in the low-energy predictions of the $A_4$ residues corresponding to nonsense wrong-signature zeros in the high-energy model. While the $t$-dependence of the $A_4$ is in good agreement with our expectations, a sign mismatch was found in the comparison between the high- and the low-energy models. The low-energy model predictions at $t \to 0$ suggest a factorizable $\rho$ contribution, while the $\omega$ exchanges indicate deviations from factorization. On the other hand, the behavior of the amplitude at $t \approx -0.5 \mbox{ GeV}^2$ suggests the very opposite. Through the use of FESR, we found that a NWSZ seems to be lacking in the $t$-channel helicity flip amplitude of the $\rho$ residue. Including this observation in our model, results in a mechanism where the dip in $\eta$ photoproduction is filled up with natural contributions, rather than genuinely assumed unnatural $b$ exchange~\cite{Gault:1971dq}. The upcoming GlueX results will be able to either confirm or refute this explanation: photon beam asymmetry measurements close to $\Sigma = +1$ within the range $-t \approx 0.5 - 0.6 \mbox{ GeV}^2$ would indicate that the absence of a dip in eta photoproduction should indeed be attributed to natural exchanges.

Inspired by the low-energy predictions, two high-energy models were presented. In the first one, we consider a conservative model with only $t$-channel exchanges that can be associated with observed meson resonances. Within the high-energy model, the $A_3$ invariant amplitude is expected to be zero, since no known mesons can contribute to it. However, the low-energy predictions suggest a large isovector $A_3$ component. Therefore, in the second model we include exchanges that correspond to, as yet, unobserved mesons. We provided predictions for the cross section and beam asymmetry at high-energies and suggested experimental signatures of factorization and novel meson exchanges.

A global analysis of low- and high-energy data of related reactions within the framework of FESR can shed light onto some of the above-mentioned inconsistencies. Especially, the inclusion of constraints from related neutral pion photoproduction amplitudes and data can resolve some of the issues. In this work, we found that the lack of dip in the cross section of $\eta$ photoproduction  is due to a dominant $A_1^\rho$ contribution, which does not have a zero in its residue. In neutral pion photoproduction, the cross section shows a dip due to a dominant $A_4^\omega$, which contains a nonsense wrong signature zero. This work, in combination with an ongoing FESR analysis in pion photoproduction.~\cite{vincent_fesr_pion} prepares the ground for such a combined analysis.

In future research, it is interesting to study whether low-energy models can provide a good description of the data when the $A_3$ invariant amplitude is forced to be small. In this respect, the FESR can be used to propagate high-energy information to constrain the low-energy models. Such an analysis is outside the scope of this work. 

All material together with an interactive website for the model will be made available on-line~\cite{JPACwebsitepaper,JPACwebsite}.

\section{Acknowledgments}
The authors would like to thank Dr.~Lothar Tiator for providing helpful comments on the manuscript.
This work was supported by the Research Foundation Flanders (FWO-Flanders). J.~Nys was supported as an `FWO-aspirant'. This material is based upon work supported in part by the U.S. Department of Energy, Office of Science, Office of Nuclear Physics under Contracts No. DE-AC05-06OR23177 and No. DE-FG0287ER40365, the National Science Foundation under Grants No. PHY-1415459 and No. NSF-PHY-1205019, and the IU Collaborative Research Grant. 
C.F.-R. is supported in part by CONACYT (Mexico) under grant No. 251817. This work was also supported by the Spanish Ministerio de Econom\'{\i}a y Competitividad (MINECO) and European FEDER funds under contract No.~FIS2014-51948-C2-2-P and SEV-2014-0398.


\appendix
\section{Kinematics and conventions}\label{sec:kinematics}
In the $s$-channel center-of-mass (c.o.m.) frame, we write the particle four-momenta as
\begin{align}
&k^\mu = (\abs{\vec{k}},\vec{k}) \,,
&q^\mu = (E_q, \vec{q}) \,, \\
&p_i^\mu = (E_i, - \vec{k})\,, 
&p_f^\mu = (E_f, -\vec{q}) \,, 
\end{align}
for which the components follow directly from the invariants 
\begin{align}
&\abs{\vec{k}} = \frac{s - M_N^2}{2\sqrt{s}} \,,
&E_q = \frac{s - M_N^2 + \mu^2}{2\sqrt{s}} \,,\\
&E_i = \frac{s + M_N^2}{2\sqrt{s}} \,,
&E_f = \frac{s + M_N^2 - \mu^2}{2\sqrt{s}} \,.
\end{align}
The c.o.m.\ energy $W$ follows from $W = \sqrt{s}$.
The eta-meson three-momentum $\vec{q}$ and c.m.\ scattering angle $\theta$ are readily determined using
\begin{align}\label{eq:costhetacm_and_q}
z_s &\equiv \cos \theta = \frac{t - u + \Delta/s}{4 \abs{\vec{k}} \abs{\vec{q}}} \,,\\
\abs{\vec{q}} &= \frac{\sqrt{(s - (M_N-\mu)^2)(s - (M_N+\mu)^2)}}{2\sqrt{s}}\,,
\end{align}
where $\Delta = M_N^2 (M_N^2 - \mu^2)$. Furthermore, we introduce
\begin{align}
t' &= t - t(z_s = +1) \,.
\end{align}
In the high-$s$ limit, $t' \rightarrow t$. We distinguish the $\pi N$ and $\eta N$ thresholds and the nucleon pole ($\nu_\pi$, $\nu_\eta$ and $\nu_N$ respectively) which can be computed using the following expressions
\begin{align}
\nu_\pi &= \frac{2 (M_N + m_\pi)^2 + t - \Sigma}{4 M_N} \label{eq:nu_pi} \,, \\
\nu_\eta &= \frac{2 (M_N + \mu)^2 + t - \Sigma}{4 M_N}= \mu + \frac{t+\mu^2}{4M_N} \,, \\
\nu_N &= \frac{2 M_N^2 + t - \Sigma}{4 M_N} =\frac{t - \mu^2}{4 M_N} \,, 
\end{align}
where $\Sigma = s + t + u = 2 M_N^2 + \mu^2$. The photon energy in the laboratory frame is given by
\begin{align}
E_\gamma^{\textrm{lab}} = \frac{s - M_N^2}{2 M_N} \,.
\end{align}

\section{Factorization}\label{sec:factorization}
In Regge theory, factorization follows from unitarity of the scattering amplitude~\cite{GellMann:1962zz,Arbab:1969zr,Gribov:1962fw,CharapSquiresFactorization}. This section details the effect of factorization on the Regge-pole amplitude. First, we derive factors which result from purely angular-momentum conservation, which must be included in the general scattering amplitude. Finally, we discuss how restrictions in the $t$-channel manifest themselves in the $s$-channel amplitudes.

In order to analytically continue the helicity amplitudes, one must identify all kinematic singularities. In Ref.~\cite{Wang:1966zza}, Wang derived the threshold, pseudo-threshold and small $\abs{t}$ factors which can lead to singularities in the parity-conserving helicity amplitudes. Once these are pulled out of the amplitude, it only contains dynamical singularities. This is for example required when the $t$-channel helicity amplitudes are Reggeized and continued for large $s$. In Ref.~\cite{Leader:1968zz} the implications of these $t$ factors on the $s$-channel amplitude were discussed. On top of that, Leader considered with rigor the effect of factorization of the residues of the $t$-channel contributions. 

For convenience of notation, let us denote the $\gamma N \rightarrow \eta N$ reaction by $1 + 2 \rightarrow 3 + 4$ with helicities $\mu_{i=1,2,3,4}$ and $\lambda_{i=1,2,3,4}$ in the $s$- and $t$-channel respectively. Let $A^{(s)}_{\mu_4 \mu_3, \mu_2 \mu_1}$ be the $s$-channel and $A^{(t)}_{\lambda_4 \lambda_2, \lambda_3 \lambda_1}$ the $t$-channel helicity amplitude~\footnote{We will explicitly denote the $s$- and $t$-channel between brackets in superscript in this section only (i.e.\ $A^{(s)}$ and $A^{(t)}$). In any other case, we consider the $s$-channel amplitudes.}. The kinematic $t$-singularities in $A^{(s)}_{\mu_4 \mu_3, \mu_2 \mu_1}$ stem entirely from the half-angle factor 
\begin{align}\label{eq:def_half_angle_sch}
\xi_{\mu \mu'}(z_s) &= \left(\frac{1+z_s}{2}\right)^{\frac{\abs{\mu+\mu'}}{2}} \left(\frac{1-z_s}{2}\right)^{\frac{\abs{\mu-\mu'}}{2}} \,,\\
\mu &= \mu_1 - \mu_2,\quad \mu' = \mu_3 - \mu_4 \nonumber \,,
\end{align}
in the rotation functions $d^J_{\mu \mu'}(z_s)$ in the partial wave expansion~\cite{Jacob:1959at}
\begin{align}\label{eq:s_ch_pw_exp}
A^{(s)}_{\mu_4 \mu_3, \mu_2 \mu_1}(s,t) &=  \sum_{J = M}^{+\infty} (2J+1)A^{(s)J}_{\mu_4 \mu_3, \mu_2 \mu_1}(s) d^J_{\mu \mu'}(z_s) \,,\\
M &= \max \{ \abs{\mu} , \abs{\mu'} \} \,. \nonumber
\end{align}
One defines the $s$-channel helicity amplitude which is free from kinematic $t$ singularities via
\begin{align}\label{eq:def_t_sing_free_sch}
\hat{A}^{(s)}_{\mu_4 \mu_3, \mu_2 \mu_1}(s,t) = A^{(s)}_{\mu_4 \mu_3, \mu_2 \mu_1}(s,t) / \xi_{\mu \mu'}(z_s) \,.
\end{align}

Since $\hat{A}^{(s)}_{\mu_4 \mu_3, \mu_2 \mu_1}$ is known to be free from $t$-singularities, and since
\begin{align}
z_s &= 1 + \frac{2 s t'}{\mathcal{S}_{12}(s) \mathcal{S}_{34}(s)} \,, \\
\mathcal{S}_{i j}^2(s) &= \left[ s - (m_i + m_j)^2 \right]\left[ s - (m_i - m_j)^2 \right] \,,
\end{align}
it is easy to see from Eq.~\eqref{eq:def_half_angle_sch} and~\eqref{eq:def_t_sing_free_sch} that the most singular behavior of $A^{(s)}_{\mu_4 \mu_3, \mu_2 \mu_1}$ is
\begin{align}\label{eq:ang_mom_behavior}
A^{(s)}_{\mu_4 \mu_3, \mu_2 \mu_1} \underset{t' \rightarrow 0}{\sim} \left( -t' \right)^{\frac{\abs{\mu - \mu'}}{2}} \underset{s\rightarrow +\infty}{\sim} (-t)^{\frac{\abs{(\mu_3 - \mu_1) - (\mu_4 - \mu_2)}}{2}} \,.
\end{align}
This behavior states that no net helicity flip is allowed at $z_s = +1$ if the angular momentum is to be conserved.
As discussed in Section~\ref{sec:helicity_amps_factorization}, the factorization of the Regge residue forces harder constraints on the small $\abs{t}$ behavior (\textit{cf}.~Eq.~\eqref{eq:t_behaviour_angular_momentum}) than would be expected from purely angular-momentum conservation (\textit{cf}.~Eq.~\eqref{eq:t_behaviour_factorization}).

In order to figure out the dominant small $\abs{t}$ dependence of the amplitudes when factorization of the $t$-channel residues is imposed, it is natural to first trace back all the $t$-factors in the $t$-channel. These results in the $t$-channel are then rotated to the $s$-channel, where the crossing matrix might introduce additional factors. Such a procedure is straightforward when there are unequal masses in both the initial and final state of the $t$-channel process~\cite{Leader:1968zz}. For the case of equal masses (such as the current one), the derivations are tedious and we will outline the general idea below. Analogously to Eq.~\eqref{eq:s_ch_pw_exp}, the $t$-channel helicity amplitude $A^{(t)}_{\lambda_4 \lambda_2, \lambda_3 \lambda_1}$ can be expanded in terms of the partial wave amplitudes $A^{(t)J}_{\lambda_4 \lambda_2, \lambda_3 \lambda_1}(t)$, and a kinematic $s$-singularity free amplitude can be defined
\begin{align}
\hat{A}^{(t)}_{\lambda_4 \lambda_2, \lambda_3 \lambda_1}(s,t) = A^{(t)}_{\lambda_4 \lambda_2, \lambda_3 \lambda_1}(s,t) \xi^{-1}_{\lambda \lambda'}(z_t) \,, \\
\lambda = \lambda_1 - \lambda_3 \,,\quad \lambda' = \lambda_2 - \lambda_4 \,. \nonumber 
\end{align}
Here, 
\begin{align}
z_t &= \frac{t^2 + t(2 s - \Sigma) + (m_1^2 - m_3^2)(m_2^2 - m_4^2)}{\mathcal{T}_{13}(t)\mathcal{T}_{24}(t)} \,, \\
\mathcal{T}_{ij}^2(t) &= \left[ t - (m_i + m_j)^2 \right]\left[ t - (m_i - m_j)^2 \right] \,. \nonumber
\end{align}
After applying the Sommerfeld-Watson transformation to the partial-wave expansion of the kinematic-singularity free, definite parity and signature amplitude, one obtains the following Regge pole contribution to the amplitude~\cite{collins}
\begin{widetext}
\begin{align}
A^{(t)}_{\lambda_4 \lambda_2, \lambda_3 \lambda_1}(s,t)& = - (-1)^{\lambda'}  (2 \alpha(t) + 1) \pi \beta_{\lambda_4 \lambda_2, \lambda_3 \lambda_1} (t) \zeta_\tau(t) d^{\alpha(t)}_{\lambda \lambda'}(z_t) \label{eq:regge_pole_representation}\,,\\
\zeta_\tau(t) &= \frac{ \tau + e^{-i\pi\alpha(t)} }{2 \sin \pi \alpha(t)} \,.
\end{align}
\end{widetext}
Assuming that the high-energy amplitude can be decomposed into a sum of Regge pole contributions Eq.~\eqref{eq:regge_pole_representation}
\begin{align}
A^{(t)}_{\lambda_4 \lambda_2, \lambda_3 \lambda_1}(s,t) = \sum_n A^{(t) n}_{\lambda_4 \lambda_2, \lambda_3 \lambda_1}(s,t) \,.
\end{align}
At leading order in $z_t$ (or equivalently $s$), the rotation functions factorize~\cite{Fox:1967zza,Leader:1968zz}
\begin{align}
&d^J_{\lambda \lambda'}(z_t) \underset{z_t \rightarrow +\infty}{\rightarrow} (-1)^{\lambda'} \mathcal{D}_{\lambda}^J(z_t) \mathcal{D}_{\lambda'}^J(z_t)\,,
\end{align}
where
\begin{widetext}
\begin{align}
\mathcal{D}_{\lambda}^J(z_t) = 
\left[(-1)^{\lambda}\left(\frac{z_t}{2}\right)^J 
\frac{\Gamma(2\alpha + 1)}{\Gamma(\alpha-\abs{\lambda} + 1)\Gamma(\alpha+\abs{\lambda} + 1)}\right]^{1/2} \,.
\end{align}
\end{widetext}
In combination with the factorization of the residues $\beta_{\lambda_4 \lambda_2, \lambda_3 \lambda_1}(t) = \beta_{\lambda_4 \lambda_2}(t) \beta_{\lambda_3 \lambda_1}(t)$~\cite{Arbab:1969zr}, the above can be cast into the factorized form
\begin{align}
A^{(t) n}_{\lambda_4 \lambda_2, \lambda_3 \lambda_1}(s,t) = - A^{(t)n}_{\lambda_4 \lambda_2}(s,t) A^{(t)n}_{\lambda_3 \lambda_1}(s,t) \,,
\end{align}
where 
\begin{align}
A^{(t)n}_{\lambda_3 \lambda_1}(s,t) &=  \left[\left(2\alpha+1\right) \pi  \zeta_\tau \right]^{1/2} \beta_{\lambda_3 \lambda_1} \mathcal{D}_{\lambda}^{\alpha}(z_t)  \,. 
\end{align}
Since the crossing matrix~\cite{Trueman:1964zzb} also factorizes
\begin{align}
R_{\lambda_4 \lambda_2, \lambda_3 \lambda_1}^{\mu_4 \mu_3, \mu_2 \mu_1} (s,t) = R_{\lambda_4 \lambda_2}^{\mu_4 \mu_2} (s,t) R_{\lambda_3 \lambda_1}^{\mu_3 \mu_1} (s,t) \,,
\end{align} 
we can write
\begin{align}
A^{(s)n}_{\mu_4 \mu_3, \mu_2 \mu_1}(s,t) = -  A^{(s)n}_{\mu_4 \mu_2}(s,t) A^{(s)n}_{\mu_3 \mu_1}(s,t) \,.
\end{align}
Hence, in the high-$s$ limit, factorization and a single Regge pole in the $t$-channel can be linked to factorization in the $s$-channel. Obviously, the behavior in Eq.~\eqref{eq:ang_mom_behavior}, is at variance with the latter. It can be shown that the simplest solution to this problem is to take~\cite{Fox:1967zza,Leader:1968zz,Cohen-Tannoudji:1968eoa}
\begin{align}
A^{(s)}_{\mu_4 \mu_3, \mu_2 \mu_1} \underset{t\rightarrow 0}{\sim} (-t)^{\frac{1}{2}\left(\abs{\mu_3 - \mu_1} + {\abs{\mu_4 - \mu_2}}\right)} \,,
\end{align}
which is obviously a more stringent constraint compared to Eq.~\eqref{eq:ang_mom_behavior}.

\section{Nucleon pole term}\label{sec:nucleon_pole_contr}
The Born contributions to the reaction amplitudes are shown diagrammatically in Fig.~\ref{fig:s_u_channel}. We decompose the contributions in the covariant basis in Eqs.~\eqref{eq:M_1}--\eqref{eq:M_4}
\begin{widetext}
\begin{align}
\begin{split}
A^{\textrm{pole}}(s,t) =& A^{s-\textrm{ch. pole}}(s,t) + A^{u-\textrm{ch. pole}}(s,t)
\\
=&-e g_{\eta NN} \overline{u}(p_f) \left[ \gamma_5 \frac{\slashed{k} + \slashed{p_i} + M_N}{s-M_N^2} \left( e_N \slashed{\epsilon} + \frac{i \kappa_N}{4 M_N}  \sigma_{\mu \nu} F^{\mu \nu} \right)  
+ \left( e_N \slashed{\epsilon} + \frac{i \kappa_N}{4M_N}  \sigma_{\mu \nu} F^{\mu \nu} \right) \frac{\slashed{p_f} - \slashed{k} + M_N}{u-M_N^2} \gamma_5 \right] u(p_i)  \\
=&\: \overline{u}(p_f) \left[e e_N g_{\eta NN} \left(\frac{1}{s-M_N^2} + \frac{1}{u-M_N^2}\right) M_1 \right.
+ 2 e e_N g_{\eta NN} \left(\frac{1}{(s-M_N^2)(u-M_N^2)}\right) M_2  \\
& -\frac{e g_{\eta NN}}{2 M_N} \kappa_N \left(\frac{1}{s-M_N^2} - \frac{1}{u-M_N^2}\right) M_3 
\left. -\frac{e g_{\eta NN}}{2 M_N} \kappa_N \left(\frac{1}{s-M_N^2} + \frac{1}{u-M_N^2}\right) M_4 \right] u(p_i)\,. 
\end{split}
\end{align}
\end{widetext}

This clearly highlights the crossing symmetry of the $A_i$. Note that $e_N = 1$ ($0$) for the proton (neutron).

\section{Subthreshold continuation}\label{sec:eta_maid_formalism}
In this section, we summarize the $\eta$-MAID 2001 formalism for resonance contributions to eta photoproduction. A resonance contribution to a multipole $\mathcal{M}_{l\pm}$ reads 

\begin{align}\label{eq:def_multipoles}
\mathcal{M}_{l \pm} (W) = &\tilde{\mathcal{M}}_{R,l \pm} \frac{m_R \Gamma_{tot}}{m_R^2 - W^2 - i m_R \Gamma_{tot}} f_{\eta N} C_{\eta N},
\end{align}
where $C_{\eta N}$ is an isospin factor and
\begin{align}
f_{\eta N} &= \zeta_{\eta N} \left[\frac{1}{(2J+1)\pi}\frac{\abs{\vec{k}} M_N}{\abs{\vec{q}} m_R}\frac{\Gamma_{\eta N}}{\Gamma_{tot}^2}\right]^{1/2} \,,\\
\Gamma_{tot} &= \Gamma_\pi + \Gamma_\eta + \Gamma_{2\pi} \,,\\
\Gamma_\pi &= \beta_{\pi N} \Gamma \left(\frac{\abs{\vec{q}_\pi}}{\abs{\vec{q}_{\pi,R}}} \right)^{2 l + 1}\left(\frac{X^2 + \abs{\vec{q}_{\pi,R}}^2}{X^2 + \abs{\vec{q}_{\pi}}^2} \right)^l \frac{m_R}{W}\,,\\
\Gamma_\eta &= \beta_{\eta N} \Gamma \left(\frac{\abs{\vec{q}}}{\abs{\vec{q}_{R}}} \right)^{2 l + 1}\left(\frac{X^2 + \abs{\vec{q}_{R}}^2}{X^2 + \abs{\vec{q}}^2} \right)^l \frac{m_R}{W} \,, \\
\Gamma_{2\pi} &= (1 - \beta_{\eta N} - \beta_{\pi N}) \Gamma \left(\frac{\abs{\vec{q}_{2\pi}}}{\abs{\vec{q}_{2\pi,R}}} \right)^{2 l + 4} \nonumber \\
&\times\left(\frac{X^2 + \abs{\vec{q}_{2\pi,R}}^2}{X^2 + \abs{\vec{q}_{2\pi}}^2} \right)^{l+2}\,.
\end{align}
The  $\zeta_{\eta N} \pm 1 $ is the relative sign between the decay of the resonance to the $\pi N$ and the $\eta N$ channels, $X$ is a scale factor related to the range of interactions responsible for the finite scattering in higher partial waves $l > 0$, and $\beta_x = \Gamma_{x}(m_R)/\Gamma$ is the branching ratio of the resonance into channel $x$. The $\vec{q}_x$ and $\vec{q}_{x,R}$ denote the center-of-mass three-momenta evaluated at $W$ and $W=m_R$ respectively. 
The parameters  $\tilde{\mathcal{M}}_{R,l\pm}$ can be related to the photo-excitation helicity amplitudes, as shown in Ref.~\cite{Chiang:2001as}.
The $C_{\eta N}$, $\tilde{\mathcal{M}}_{R,l\pm}$, $\Gamma$, $m_R$, $\beta_{\pi N}$ and $\beta_{\eta N}$ were obtained in a fit to the world data in Ref.~\cite{Chiang:2001as} for the proton and
Ref.~\cite{Fix:2007st} for the neutron.
For the subthreshold evaluation of the multipoles, we take $\abs{\vec{q}_x} = \Re{ \left[ {\sqrt{(s - (M_N-m_x)^2)(s - (M_N+m_x)^2)}}/{(2\sqrt{s})} \right]}$ in the evaluation of the energy dependent decay widths.

Finally, the CGLN  amplitudes, $\mathcal{F}_i$ defined in Eq.~\eqref{eq:def_CGLN_Fi} are constructed from the multipoles using the relations summarized in Eqs.~\eqref{eq:multipole_decomposition_a}--\eqref{eq:multipole_decomposition_d}.

\color{black}

\section{$b_1$ couplings}\label{sec:b_coupling}
Using vector-meson dominance (VMD), SU(3) flavor symmetry and the OZI rule, one obtains the following relations between the neutral pseudoscalar-meson photoproduction amplitudes
\begin{align}
A(\eta) = \sqrt{3}\mathcal{A} &\left[ A_\rho(\pi^0) + A_b(\pi^0) \right. \nonumber\\
&\left.+ \frac{1}{9} (A_\omega(\pi^0) + A_h(\pi^0))\right] \label{eq:gault_relation}\,,
\end{align}
where the $\sqrt{3}$ is related to ideal $\omega-\phi$ mixing and $\mathcal{A}=1.55$ to $\eta-\eta'$ mixing~\cite{Worden:1972dc,Dalitz:1965ey,Martin:1972pt}. The $b_1$ decays dominantly through $b_1 \rightarrow \pi^0 \omega$. In Ref.~\cite{Haglin:1994yv} it is shown that $g_{b_1 \pi \omega} = 9.77$ from the corresponding decay width. Using VMD, one can relate $g_{b_1 \pi \omega}$ to $g_{b_1 \pi \gamma}$
\begin{align}
g_{b_1 \pi \gamma} = \frac{e}{f_\omega} g_{b_1 \pi \omega} = 0.189 \,,
\end{align}
where $f_\omega = 3 f_\rho$ is the universal coupling constant of the $\omega$ meson. This constant is not well constrained. We take $f_\rho = 5.2$ as in Ref.~\cite{Yu:2011zu}. After applying Eq.~\eqref{eq:gault_relation}, one obtains $g_{b_1 \eta \gamma} = 0.51$ . The nucleon vertex $b_1 NN$ can be estimated through axial-vector meson dominance (AVMD) as demonstrated in Ref.~\cite{Yu:2011zu}. Yu \textit{et al.}~\cite{Yu:2011zu} have illustrated that their estimate of $g^t_{b_1 NN} = -14$ is in good agreement with more fundamental theories (see Table~4 in Ref.~\cite{Yu:2011zu} and the discussion thereof). Finally, in our notation, we obtain
\begin{align}
g_2^b = g_{b_1 \eta \gamma} \frac{g^t_{b_1 NN}}{2 M_N}= -3.8\mbox{ GeV}^{-2} \,.
\end{align}

\section{Amplitude bases}\label{sec:amplitude_bases}
The invariant amplitudes are defined in Eqs~\eqref{eq:M_1}--\eqref{eq:M_4}. Here we summarize their relation to the $s$-channel electric and magnetic multipoles~\cite{CGLNpaper}. In  terms of spinor amplitudes, 
\begin{align}
A = \frac{4\pi W}{M_N} \chi_f^\dag \mathcal{F} \chi_i \, ,
\end{align}
where $\chi_i$ ($\chi_f$) is the initial (final) nucleon Pauli spinor in the center-of-mass frame and
\begin{align}\label{eq:def_CGLN_Fi}
\mathcal{F} = & \sigmavec \cdot \epsilonvec \mathcal{F}_1 -i \sigmavec\cdot\hat{\vec{q}} \sigmavec\cdot(\hat{\vec{k}} \times \epsilonvec) \mathcal{F}_2 \nonumber \\
&+ \sigmavec\cdot\hat{\vec{k}} \hat{\vec{q}} \cdot \epsilonvec \mathcal{F}_3 + \sigmavec \cdot \hat{\vec{q}} \hat{\vec{q}} \cdot \epsilonvec \mathcal{F}_4 \, ,
\end{align}
where, $\hat{\vec{q}} = \vec{q}/\abs{\vec{q}}$ and $\hat{\vec{k}} = \vec{k}/\abs{\vec{k}}$.
The amplitudes $\mathcal{F}_i$ are given in terms of the multipoles $\mathcal{M}_{l \pm}$ by, 
\begin{align}
\mathcal{F}_1 &= \sum\limits_{l = 0} \left( E_{l+} + l M_{l+} \right)P'_{l+1} +  \left( E_{l-} + (l+1) M_{l-} \right)P'_{l-1} \,, \label{eq:multipole_decomposition_a}\\
\mathcal{F}_2 &= \sum\limits_{l = 1} \left( (l+1) M_{l+} + l M_{l-} \right)P'_{l}\,,\\
\mathcal{F}_3 &= \sum\limits_{l = 1} \left( E_{l+} - l M_{l+} \right)P''_{l+1} +  \left( E_{l-} + M_{l-} \right)P''_{l-1} \,,\\
\mathcal{F}_4 &= \sum\limits_{l = 2} \left( - E_{l+} + M_{l+} - E_{l-} - M_{l-} \right)P''_{l}\,. \label{eq:multipole_decomposition_d}
\end{align}
The derivatives of the Legendre polynomials ($P^{(n)}_l$) are a function of $\cos\theta$, while the multipoles depend on $s$ only.
The invariant amplitudes $A_i$ are obtained from the $\mathcal{F}_i$'s using
\begin{align}
A_1 =& \mathcal{N} \left[ \frac{W+M_N}{W-M_N} \tilde{\mathcal{F}}_1 - (E_f + M_N)\tilde{\mathcal{F}}_2 \right.\nonumber\\
&\left. + M_N \frac{t-\mu^2}{(W-M_N)^2} \tilde{\mathcal{F}}_3 + M_N \frac{(E_f + M_N)(t-\mu^2)}{W^2 - M_N^2} \tilde{\mathcal{F}}_4
 \right] \,, \label{eq:A1_ifo_Fi}\\
A_2 =&\frac{\mathcal{N}}{W-M_N} \left[ \tilde{\mathcal{F}}_3 - (E_f + M_N) \tilde{\mathcal{F}}_4 \right] \,, \\
A_3 =&\frac{\mathcal{N}}{W-M_N} \left[ \tilde{\mathcal{F}}_1 + (E_f + M_N)\tilde{\mathcal{F}}_2 \right. \nonumber \\
&+\left(W+M_N + \frac{t-\mu^2}{2 (W-M_N)}\right) \tilde{\mathcal{F}}_3 \nonumber \\
&\left.+\left( W-M_N + \frac{t-\mu^2}{2(W+M_N)} \right)(E_f + M_N) \tilde{\mathcal{F}}_4 \right] \,, \\
A_4 =&\frac{\mathcal{N}}{W-M_N} \left[ \tilde{\mathcal{F}}_1 + (E_f + M_N)\tilde{\mathcal{F}}_2 \right. \nonumber \\
&\left. + \frac{t-\mu^2}{2 (W-M_N)} \tilde{\mathcal{F}}_3 + \frac{t-\mu^2}{2(W+M_N)} (E_f + M_N) \tilde{\mathcal{F}}_4 \right]\label{eq:A4_ifo_Fi} \,,
\end{align}
where $\mathcal{N} = 4 \pi /\sqrt{(E_i + M_N) (E_f + M_N)}$ and the reduced Chew-Goldberger-Low-Nambu (CGLN) amplitudes are defined by
\begin{align}
&\tilde{\mathcal{F}}_1 = \mathcal{F}_1\,,  &\tilde{\mathcal{F}}_2 = \mathcal{F}_2/\abs{\vec{q}} \,, \nonumber \\
&\tilde{\mathcal{F}}_3 = \mathcal{F}_3/\abs{\vec{q}} \,, &\tilde{\mathcal{F}}_4 = \mathcal{F}_4/\abs{\vec{q}}^2 \,.
\end{align}
The factors of $\abs{\vec{q}}$ remove the kinematic threshold zeros that appear in the multipole decomposition of $\mathcal{F}_i$, $\mathcal{M}_{l\pm} \underset{\abs{\vec{q}}\rightarrow 0}{\sim} \abs{\vec{q}}^l$.
Explicitly, the reduced CGLN amplitudes up to and including $D$-waves ($l=2$) are
\begin{align}
\tilde{F_1} =& E_{0+}+ E_{2-} + 3 M_{2-} + 3  (E_{1+} + M_{1+}) \cos \theta  \nonumber \\
+& 3/2(E_{2+} + 2 M_{2+})(5 \cos^2\theta - 1)\,, \\
\tilde{F_2} =& \left[ M_{1-} + 2 M_{1+} + 3  (2 M_{2-} + 3 M_{2+})\cos \theta \right]/\abs{\vec{q}}\,, \\
\tilde{F_3} =& \left[ 3 (E_{1+} - M_{1+}) + 15  (E_{2+} - M_{2+})\cos \theta\right]/\abs{\vec{q}}\,, \\
\tilde{F_4} =&  3\left[-E_{2-} - E_{2+} - M_{2-} + M_{2+}\right]/\abs{\vec{q}}^2\,.
\end{align}

\bibliography{biblio.bib}
\end{document}